\documentclass[conference]{IEEEtran}

\IEEEoverridecommandlockouts

\usepackage{cite}
\usepackage{amsmath,amssymb,amsfonts}
\usepackage{graphicx}
\usepackage{textcomp}
\usepackage{xcolor}
\usepackage{soul}

\usepackage{algorithmic}
\usepackage[algo2e,ruled,vlined,linesnumbered]{algorithm2e}
\DontPrintSemicolon
\SetAlgoVlined
\SetKwProg{Fn}{Function}{}{end}
\SetKwFor{ParallelDo}{parallel do}{}{}
\SetKwFor{ParallelFor}{for}{do parallel}{}

\usepackage{numprint}
\npthousandsep{\,}
\npdecimalsign{.}

\def\BibTeX{{\rm B\kern-.05em{\sc i\kern-.025em b}\kern-.08em
T\kern-.1667em\lower.7ex\hbox{E}\kern-.125emX}}

\newif\ifshowcomments

\newif\ifsubmission

\usepackage{tikz}
\usepackage{amsbsy}
\usepackage{mathtools}
\usepackage{wasysym}
\usepackage{tabularx}
\usepackage{pifont} 
\usepackage{booktabs}

\usepackage{capt-of}
\usepackage{multirow} 

\usepackage[hidelinks]{hyperref}
\usepackage{cleveref}

\newcommand{\symbInfeasible}{\ensuremath{\times}}

\definecolor{fuchsiapink}{rgb}{1.0, 0.47, 1.0}
\definecolor{darkgreen}{rgb}{0.0, 0.7, 0.2}

\ifshowcomments
  \newcommand{\tobi}[1]{{\color{fuchsiapink}[TH: #1]}}
  \newcommand{\lars}[1]{{\color{cyan}{[Lars: #1]}}}
  \newcommand{\peter}[1]{{\color{blue}[PS: #1]}}
  \newcommand{\psfrage}[1]{{\color{blue}[PS: #1]}}
  \newcommand{\daniel}[1]{{\color{orange}[DSee: #1]}}
  \newcommand{\danielsal}[1]{{\color{darkgreen}[DSal: #1]}}
  \newcommand{\nikolai}[1]{{\color{darkgreen}[NM: #1]}}
  \newcommand{\todo}[1]{{\textcolor{red}{\bf [TODO]} \emph{#1}}}

\else
  \newcommand{\tobi}[1]{}
  \newcommand{\lars}[1]{}
  \newcommand{\peter}[1]{}
  \newcommand{\psfrage}[1]{}
  \newcommand{\daniel}[1]{}
  \newcommand{\danielsal}[1]{}
  \newcommand{\nikolai}[1]{}
  \newcommand{\todo}[1]{}

\fi

\newcommand{\etal}{{et al}.}

\newcommand{\xmark}{\ding{55}}

\newcommand{\pluseq}{\mathrel{+}=}
\newcommand{\minuseq}{\mathrel{-}=}

\newcommand{\Partition}{\ensuremath{{\Pi}}}%
\DeclareMathAlphabet{\mathpzc}{OT1}{pzc}{m}{n}

\newcommand{\splitatcommas}[1]{%
\begingroup
\begingroup\lccode`~=`, \lowercase{\endgroup
\edef~{\mathchar\the\mathcode`, \penalty0 \noexpand\hspace{0pt plus 1em}}%
}\mathcode`,="8000 #1%
\endgroup
}

\newcommand{\gpp}{\texttt{g++}}

\newcommand{\Instance}[1]{\textsf{#1}}

\newcommand{\Partitioner}[1]{\textsc{#1}} 
\newcommand{\KaMinPar}{\Partitioner{KaMinPar}\xspace}
\newcommand{\XtraPuLP}{\Partitioner{XtraPuLP}\xspace}
\newcommand{\dKaMinPar}{\Partitioner{dKaMinPar}\xspace}
\newcommand{\ParMetis}{\Partitioner{ParMetis}\xspace}
\newcommand{\MtMetis}{\Partitioner{Mt-Metis}\xspace}
\newcommand{\TeraPart}{\Partitioner{TeraPart}\xspace}
\newcommand{\xTeraPart}{\Partitioner{xTeraPart}\xspace}
\newcommand{\SEM}{\Partitioner{SEM}\xspace}
\newcommand{\HeiStream}{\Partitioner{HeiStream}\xspace}

\ifsubmission
\newcommand{\HoreKa}{$\langle$\emph{redacted}$\rangle$\xspace}
\else
\newcommand{\HoreKa}{HoreKa\xspace}
\fi

\DeclareMathOperator*{\argmax}{arg\,max}
\newcommand{\Rvalue}[1]{#1}

\DeclareMathOperator{\nc}{nc}
\DeclareMathOperator{\cut}{cut}

\newcommand{\RnumGraphsInSetA}{\Rvalue{\numprint{72}}}

\newcommand{\RsmallestNumEdgesInSetAMillion}{\Rvalue{\numprint{5.4}}}

\newcommand{\RlargestNumEdgesInSetABillion}{\Rvalue{\numprint{1.8}}}

\newcommand{\RioTimeUncompressedParS}{\Rvalue{\numprint{177}}}
\newcommand{\RioTimeUncompressedSeqS}{\Rvalue{\numprint{572}}}
\newcommand{\RioTimeCompressedParS}{\Rvalue{\numprint{179}}}
\newcommand{\RioTimeCompressedSeqS}{\Rvalue{\numprint{2905}}}
\newcommand{\RufmLpSpHeapGiB}{\Rvalue{\numprint{3.45}}}
\newcommand{\RufmLpTpHeapGiB}{\Rvalue{\numprint{2.58}}}

\newcommand{\RufmLpUcHeapGiB}{\Rvalue{\numprint{1.79}}}
\newcommand{\RufmLpTpVsSpHeapPerc}{\Rvalue{\numprint{25.3}}}
\newcommand{\RufmLpGcVsTpHeapPerc}{\Rvalue{\numprint{24.2}}}
\newcommand{\RufmLpUcVsGcHeapPerc}{\Rvalue{\numprint{8.4}}}
\newcommand{\RufmLpTeraPartVsKaMinParHeapPerc}{\Rvalue{\numprint{48.1}}}
\newcommand{\RufmLpSpTimeS}{\Rvalue{\numprint{1.59}}}
\newcommand{\RufmLpTpTimeS}{\Rvalue{\numprint{1.40}}}

\newcommand{\RufmLpUcTimeS}{\Rvalue{\numprint{1.48}}}
\newcommand{\RufmLpTpVsSpTimePerc}{\Rvalue{\numprint{11.8}}}
\newcommand{\RufmLpTpVsGcTimePerc}{\Rvalue{\numprint{6.0}}}
\newcommand{\RufmLpUcVsGcTimePerc}{\Rvalue{\numprint{0.6}}}
\newcommand{\RufmLpTeraPartVsKaMinParTimePerc}{\Rvalue{\numprint{6.7}}}

\newcommand{\RufmLpTpVsSpTimeLargeDegPerc}{\Rvalue{\numprint{23.2}}}
\newcommand{\RufmLpNumInstances}{\Rvalue{\numprint{504}}}
\newcommand{\RufmLpNumLargeInstances}{\Rvalue{\numprint{189}}}
\newcommand{\RufmLpNumVeryLargeInstances}{\Rvalue{\numprint{28}}}
\newcommand{\RufmLpLargeTeraPartVsKaMinParHeapPerc}{\Rvalue{\numprint{63.7}}}
\newcommand{\RufmLpVeryLargeTeraPartVsKaMinParHeapPerc}{\Rvalue{\numprint{81.1}}}
\newcommand{\RufmLpKaMinParVsTeraPartCutPerc}{\Rvalue{\numprint{0.03}}}

\newcommand{\RufmLpMtMetisNumImbalancedInstances}{\Rvalue{\numprint{320}}}
\newcommand{\RufmLpMtMetisVsKaMinParTimeRatio}{\Rvalue{\numprint{3.9}}}
\newcommand{\RufmLpMtMetisVsTeraPartTimeRatio}{\Rvalue{\numprint{4.2}}}
\newcommand{\RufmLpMtMetisVsKaMinParRSSRatio}{\Rvalue{\numprint{2.7}}}
\newcommand{\RufmLpMtMetisVsTeraPartRSSRatio}{\Rvalue{\numprint{4.4}}}
\newcommand{\RufmCompressionRatio}{\Rvalue{\numprint{3.2}}}
\newcommand{\RufmCompressionRatioFiniteElement}{\Rvalue{\numprint{5.7}}}
\newcommand{\RufmLpSpeedupNumInstances}{\Rvalue{\numprint{144}}}
\newcommand{\RufmLpSpeedupNumLargeInstances}{\Rvalue{\numprint{15}}}
\newcommand{\RufmLpSpeedupNumMediumInstances}{\Rvalue{\numprint{48}}}
\newcommand{\RufmLpSpeedupTwelve}{\Rvalue{\numprint{8.7}}}
\newcommand{\RufmLpSpeedupTwelveMedium}{\Rvalue{\numprint{10.2}}}

\newcommand{\RufmLpSpeedupTwentyFour}{\Rvalue{\numprint{13.0}}}
\newcommand{\RufmLpSpeedupTwentyFourMedium}{\Rvalue{\numprint{17.0}}}

\newcommand{\RufmLpSpeedupFortyEight}{\Rvalue{\numprint{16.5}}}
\newcommand{\RufmLpSpeedupFortyEightMedium}{\Rvalue{\numprint{24.7}}}

\newcommand{\RufmLpSpeedupNinetySix}{\Rvalue{\numprint{17.3}}}
\newcommand{\RufmLpSpeedupNinetySixMedium}{\Rvalue{\numprint{29.8}}}
\newcommand{\RufmLpSpeedupNinetySixLarge}{\Rvalue{\numprint{41.6}}}
\newcommand{\RtegHeiStreamVsTeraPartRggCutRatio}{\Rvalue{\numprint{3.1}}}
\newcommand{\RtegHeiStreamVsTeraPartRhgCutRatio}{\Rvalue{\numprint{14.8}}}

\newcommand{\RhwgMinCompressionRatioRounded}{\Rvalue{\numprint{5}}}

\newcommand{\RhwgMaxCompressionRatioRounded}{\Rvalue{\numprint{11}}}
\newcommand{\RhwgMinCompressionRatioGapOnly}{\Rvalue{\numprint{2.7}}}
\newcommand{\RhwgMaxCompressionRatioGapOnly}{\Rvalue{\numprint{3.4}}}
\newcommand{\RhwgEuSpVsTpTimeRatio}{\Rvalue{\numprint{4.6}}}
\newcommand{\RhwgEuSpVsUcTimeRatio}{\Rvalue{\numprint{3.1}}}
\newcommand{\RhwgCluewebKaMinParVsTeraPartHeapRatio}{\Rvalue{\numprint{12.5}}}
\newcommand{\RhwgEuKaMinParVsTeraPartHeapRatio}{\Rvalue{\numprint{15.7}}}
\newcommand{\RhwgGshKaMinParVsTeraPartHeapRatio}{\Rvalue{\numprint{12.9}}}
\newcommand{\RhwgHyperlinkMinHeapGiB}{\Rvalue{\numprint{279}}}
\newcommand{\RhwgHyperlinkMaxHeapGiB}{\Rvalue{\numprint{306}}}
\newcommand{\RhwgUkKaMinParVsTeraPartHeapRatio}{\Rvalue{\numprint{15.7}}}

\newcommand{\RhwgGcVsUcLargeKHeapPerc}{\Rvalue{\numprint{40.7}}}
\newcommand{\RhwgGcVsUcSmallKHeapPerc}{\Rvalue{\numprint{7.5}}}
\newcommand{\RufmFmFullVsHashingHeapRatio}{\Rvalue{\numprint{2.7}}}
\newcommand{\RufmFmFullVsHashingHeapRatioMedium}{\Rvalue{\numprint{5.8}}}
\newcommand{\RufmFmFullVsHashingTimePerc}{\Rvalue{\numprint{1.6}}}
\newcommand{\RufmFmFullVsHashingTimePercMedium}{\Rvalue{\numprint{0.03}}}
\newcommand{\RufmFmKmerLargeKHeapGiB}{\Rvalue{\numprint{14.7}}}
\newcommand{\RufmFmHashingVsOnTheFlyTimeRatio}{\Rvalue{\numprint{2.7}}}
\newcommand{\RufmFmHashingVsOnTheFlyNumOrderOfMagnitudeSpeedup}{\Rvalue{\numprint{67}}}
\newcommand{\RufmFmOnTheFlyTimeouts}{\Rvalue{\numprint{16}}}

\usepackage{tikz}
\usetikzlibrary{external}
\usetikzlibrary{calc}
\makeatletter
\newcommand{\linebreakand}{%
  \end{@IEEEauthorhalign}
  \hfill\mbox{}\par
  \mbox{}\hfill\begin{@IEEEauthorhalign}
}
\makeatother

\begin{document}

\title{Tera-Scale Multilevel Graph Partitioning}

\ifsubmission
\else
\author{
  \IEEEauthorblockN{Daniel Salwasser}
  \IEEEauthorblockA{\textit{Karlsruhe Institute of Technology} \\
    daniel.salwasser@student.kit.edu}
  \and
  \IEEEauthorblockN{Daniel Seemaier}
  \IEEEauthorblockA{\textit{Karlsruhe Institute of Technology} \\
    daniel.seemaier@kit.edu}
  \linebreakand
  \IEEEauthorblockN{Lars Gottesbüren}
  \IEEEauthorblockA{\textit{Google Research, Zürich} \\
    gottesbueren@google.com}
  \and
  \IEEEauthorblockN{Peter Sanders}
  \IEEEauthorblockA{\textit{Karlsruhe Institute of Technology} \\
    sanders@kit.edu}
}
\fi

\maketitle

\begin{abstract}
  We present TeraPart, a memory-efficient multilevel graph partitioning method that is designed to scale to extremely large graphs.
  In balanced graph partitioning, the goal is to divide the vertices into $k$ blocks with balanced size while cutting as few edges as possible.
  Due to its NP-hard nature, heuristics are prevalent in this field, with the multilevel framework as the state-of-the-art method.
  Recent work has seen tremendous progress in speeding up partitioning algorithms through parallelism. The current obstacle in scaling to larger graphs is the high memory usage due to auxiliary data structures and storing the graph itself in memory.
  In this paper, we present and study several optimizations to significantly reduce their memory footprint.

  We devise parallel label propagation clustering and graph contraction algorithms that use $O(n)$ auxiliary space instead of $O(np)$, where $p$ is the number of processors. 
  Moreover, we employ an existing compressed graph representation that enables iterating over a neighborhood by on-the-fly decoding at speeds close to the uncompressed graph.
  Combining these optimizations yields up to a 16-fold reduction in peak memory, while retaining the same solution quality and similar speed.
  This configuration can partition a graph with \emph{one trillion} edges  in under 8 minutes \emph{on a single machine} using around 900\,GiB of RAM.
  This is the first work to employ the multilevel framework at this scale, which is vital to achieving low edge cuts.
  Moreover, our distributed memory implementation handles graphs of up to 16 trillion edges on 128 machines with 256\,GiB each in just under 10 minutes.
    Finally, we present a version of shared-memory parallel FM local search that uses $O(m)$ space instead of $O(nk)$, reducing peak memory by factor \RufmFmFullVsHashingHeapRatioMedium{} on medium-sized graphs without affecting running time.
\end{abstract}
%

\section{Introduction}

 \peter{I am missing a section on the distributed approach which looks like a major contribution. Below I also make a few shortening suggestions.}
 Balanced graph partitioning is a central and well-studied problem in computer science with a wide
 range of applications such as minimizing communication under load-balanced workloads in distributed
 databases, graph processing, or scientific computing simulations. The task is to divide the
 vertices $V$ of a graph $G=(V,E)$ into $k \in \mathbb{N}$ disjoint blocks $V_1, \dots, V_k
   \subseteq V$ of roughly equal size $|V_i| \leq (1+\varepsilon)\frac{|V|}{k} $ (the balance
 constraint) while minimizing the number of edges connecting different blocks $\sum_{i < j}^k |\{ \{
   u, v \} \in E \mid u \in V_i, v \in V_j \}|$ (the edge cut).
 
 Most state-of-the-art graph partitioning methods employ the multilevel scheme, which follows the
 three phases of \emph{coarsening}, \emph{initial partitioning} and \emph{uncoarsening}. In the
 coarsening phase, we iteratively reduce the graph size by contracting clusters of similar vertices,
 until the graph is sufficiently small to run a portfolio of initial partitioning algorithms. In the
 uncoarsening phase, we revert the contractions in reverse order, and project the solution up to the
 next larger graph. On each contraction level, we employ local search algorithms such as
 FM~\cite{FM} or label propagation~\cite{LABEL_PROPAGATION} to improve the solution.

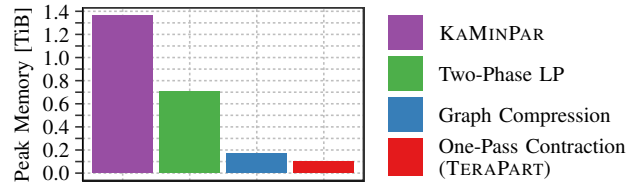
\begin{figure}
   \centering
\begin{tikzpicture}[x=1pt,y=1pt]
\definecolor{fillColor}{RGB}{255,255,255}
\path[use as bounding box,fill=fillColor,fill opacity=0.00] (0,0) rectangle (252.94, 79.50);
\begin{scope}
\path[clip] (  0.00,  0.00) rectangle (252.94, 79.50);
\definecolor{drawColor}{RGB}{255,255,255}
\definecolor{fillColor}{RGB}{255,255,255}

\path[draw=drawColor,line width= 0.6pt,line join=round,line cap=round,fill=fillColor] (  0.00,  0.00) rectangle (252.94, 79.50);
\end{scope}
\begin{scope}
\path[clip] ( 30.48,  8.25) rectangle (137.00, 74.00);
\definecolor{fillColor}{RGB}{255,255,255}

\path[fill=fillColor] ( 30.48,  8.25) rectangle (137.00, 74.00);
\definecolor{drawColor}{RGB}{190,190,190}

\path[draw=drawColor,line width= 0.6pt,dash pattern=on 1pt off 1pt ,line join=round] ( 30.48, 11.24) --
	(137.00, 11.24);

\path[draw=drawColor,line width= 0.6pt,dash pattern=on 1pt off 1pt ,line join=round] ( 30.48, 15.61) --
	(137.00, 15.61);

\path[draw=drawColor,line width= 0.6pt,dash pattern=on 1pt off 1pt ,line join=round] ( 30.48, 19.99) --
	(137.00, 19.99);

\path[draw=drawColor,line width= 0.6pt,dash pattern=on 1pt off 1pt ,line join=round] ( 30.48, 24.37) --
	(137.00, 24.37);

\path[draw=drawColor,line width= 0.6pt,dash pattern=on 1pt off 1pt ,line join=round] ( 30.48, 28.74) --
	(137.00, 28.74);

\path[draw=drawColor,line width= 0.6pt,dash pattern=on 1pt off 1pt ,line join=round] ( 30.48, 33.12) --
	(137.00, 33.12);

\path[draw=drawColor,line width= 0.6pt,dash pattern=on 1pt off 1pt ,line join=round] ( 30.48, 37.49) --
	(137.00, 37.49);

\path[draw=drawColor,line width= 0.6pt,dash pattern=on 1pt off 1pt ,line join=round] ( 30.48, 41.87) --
	(137.00, 41.87);

\path[draw=drawColor,line width= 0.6pt,dash pattern=on 1pt off 1pt ,line join=round] ( 30.48, 46.24) --
	(137.00, 46.24);

\path[draw=drawColor,line width= 0.6pt,dash pattern=on 1pt off 1pt ,line join=round] ( 30.48, 50.62) --
	(137.00, 50.62);

\path[draw=drawColor,line width= 0.6pt,dash pattern=on 1pt off 1pt ,line join=round] ( 30.48, 54.99) --
	(137.00, 54.99);

\path[draw=drawColor,line width= 0.6pt,dash pattern=on 1pt off 1pt ,line join=round] ( 30.48, 59.37) --
	(137.00, 59.37);

\path[draw=drawColor,line width= 0.6pt,dash pattern=on 1pt off 1pt ,line join=round] ( 30.48, 63.75) --
	(137.00, 63.75);

\path[draw=drawColor,line width= 0.6pt,dash pattern=on 1pt off 1pt ,line join=round] ( 30.48, 68.12) --
	(137.00, 68.12);

\path[draw=drawColor,line width= 0.6pt,dash pattern=on 1pt off 1pt ,line join=round] ( 30.48, 72.50) --
	(137.00, 72.50);

\path[draw=drawColor,line width= 0.6pt,dash pattern=on 1pt off 1pt ,line join=round] ( 45.70,  8.25) --
	( 45.70, 74.00);

\path[draw=drawColor,line width= 0.6pt,dash pattern=on 1pt off 1pt ,line join=round] ( 71.06,  8.25) --
	( 71.06, 74.00);

\path[draw=drawColor,line width= 0.6pt,dash pattern=on 1pt off 1pt ,line join=round] ( 96.42,  8.25) --
	( 96.42, 74.00);

\path[draw=drawColor,line width= 0.6pt,dash pattern=on 1pt off 1pt ,line join=round] (121.78,  8.25) --
	(121.78, 74.00);
\definecolor{fillColor}{RGB}{228,26,28}

\path[fill=fillColor] (110.37, 11.24) rectangle (133.19, 15.61);
\definecolor{fillColor}{RGB}{77,175,74}

\path[fill=fillColor] ( 59.65, 11.24) rectangle ( 82.47, 42.28);
\definecolor{fillColor}{RGB}{152,78,163}

\path[fill=fillColor] ( 34.29, 11.24) rectangle ( 57.11, 71.01);
\definecolor{fillColor}{RGB}{55,126,184}

\path[fill=fillColor] ( 85.01, 11.24) rectangle (107.83, 18.60);
\definecolor{drawColor}{gray}{0.20}

\path[draw=drawColor,line width= 1.1pt,line join=round,line cap=round] ( 30.48,  8.25) rectangle (137.00, 74.00);
\end{scope}
\begin{scope}
\path[clip] (  0.00,  0.00) rectangle (252.94, 79.50);
\definecolor{drawColor}{RGB}{0,0,0}

\path[draw=drawColor,line width= 0.2pt,line join=round] ( 30.48,  8.25) --
	( 30.48, 74.00);
\end{scope}
\begin{scope}
\path[clip] (  0.00,  0.00) rectangle (252.94, 79.50);
\definecolor{drawColor}{RGB}{0,0,0}

\node[text=drawColor,anchor=base east,inner sep=0pt, outer sep=0pt, scale=  0.80] at ( 25.53,  8.48) {0.0};

\node[text=drawColor,anchor=base east,inner sep=0pt, outer sep=0pt, scale=  0.80] at ( 25.53, 17.23) {0.2};

\node[text=drawColor,anchor=base east,inner sep=0pt, outer sep=0pt, scale=  0.80] at ( 25.53, 25.99) {0.4};

\node[text=drawColor,anchor=base east,inner sep=0pt, outer sep=0pt, scale=  0.80] at ( 25.53, 34.74) {0.6};

\node[text=drawColor,anchor=base east,inner sep=0pt, outer sep=0pt, scale=  0.80] at ( 25.53, 43.49) {0.8};

\node[text=drawColor,anchor=base east,inner sep=0pt, outer sep=0pt, scale=  0.80] at ( 25.53, 52.24) {1.0};

\node[text=drawColor,anchor=base east,inner sep=0pt, outer sep=0pt, scale=  0.80] at ( 25.53, 60.99) {1.2};

\node[text=drawColor,anchor=base east,inner sep=0pt, outer sep=0pt, scale=  0.80] at ( 25.53, 69.74) {1.4};
\end{scope}
\begin{scope}
\path[clip] (  0.00,  0.00) rectangle (252.94, 79.50);
\definecolor{drawColor}{gray}{0.20}

\path[draw=drawColor,line width= 0.6pt,line join=round] ( 27.73, 11.24) --
	( 30.48, 11.24);

\path[draw=drawColor,line width= 0.6pt,line join=round] ( 27.73, 15.61) --
	( 30.48, 15.61);

\path[draw=drawColor,line width= 0.6pt,line join=round] ( 27.73, 19.99) --
	( 30.48, 19.99);

\path[draw=drawColor,line width= 0.6pt,line join=round] ( 27.73, 24.37) --
	( 30.48, 24.37);

\path[draw=drawColor,line width= 0.6pt,line join=round] ( 27.73, 28.74) --
	( 30.48, 28.74);

\path[draw=drawColor,line width= 0.6pt,line join=round] ( 27.73, 33.12) --
	( 30.48, 33.12);

\path[draw=drawColor,line width= 0.6pt,line join=round] ( 27.73, 37.49) --
	( 30.48, 37.49);

\path[draw=drawColor,line width= 0.6pt,line join=round] ( 27.73, 41.87) --
	( 30.48, 41.87);

\path[draw=drawColor,line width= 0.6pt,line join=round] ( 27.73, 46.24) --
	( 30.48, 46.24);

\path[draw=drawColor,line width= 0.6pt,line join=round] ( 27.73, 50.62) --
	( 30.48, 50.62);

\path[draw=drawColor,line width= 0.6pt,line join=round] ( 27.73, 54.99) --
	( 30.48, 54.99);

\path[draw=drawColor,line width= 0.6pt,line join=round] ( 27.73, 59.37) --
	( 30.48, 59.37);

\path[draw=drawColor,line width= 0.6pt,line join=round] ( 27.73, 63.75) --
	( 30.48, 63.75);

\path[draw=drawColor,line width= 0.6pt,line join=round] ( 27.73, 68.12) --
	( 30.48, 68.12);

\path[draw=drawColor,line width= 0.6pt,line join=round] ( 27.73, 72.50) --
	( 30.48, 72.50);
\end{scope}
\begin{scope}
\path[clip] (  0.00,  0.00) rectangle (252.94, 79.50);
\definecolor{drawColor}{RGB}{0,0,0}

\path[draw=drawColor,line width= 0.2pt,line join=round] ( 30.48,  8.25) --
	(137.00,  8.25);
\end{scope}
\begin{scope}
\path[clip] (  0.00,  0.00) rectangle (252.94, 79.50);
\definecolor{drawColor}{RGB}{0,0,0}

\node[text=drawColor,rotate= 90.00,anchor=base,inner sep=0pt, outer sep=0pt, scale=  0.80] at ( 10.23, 41.12) {Peak Memory [TiB]};
\end{scope}
\begin{scope}
\path[clip] (  0.00,  0.00) rectangle (252.94, 79.50);
\definecolor{fillColor}{RGB}{255,255,255}

\path[fill=fillColor] (145.34, 57.20) rectangle (159.80, 71.66);
\end{scope}
\begin{scope}
\path[clip] (  0.00,  0.00) rectangle (252.94, 79.50);
\definecolor{fillColor}{RGB}{152,78,163}

\path[fill=fillColor] (146.05, 57.91) rectangle (159.09, 70.95);
\end{scope}
\begin{scope}
\path[clip] (  0.00,  0.00) rectangle (252.94, 79.50);
\definecolor{fillColor}{RGB}{255,255,255}

\path[fill=fillColor] (145.34, 41.75) rectangle (159.80, 56.20);
\end{scope}
\begin{scope}
\path[clip] (  0.00,  0.00) rectangle (252.94, 79.50);
\definecolor{fillColor}{RGB}{77,175,74}

\path[fill=fillColor] (146.05, 42.46) rectangle (159.09, 55.49);
\end{scope}
\begin{scope}
\path[clip] (  0.00,  0.00) rectangle (252.94, 79.50);
\definecolor{fillColor}{RGB}{255,255,255}

\path[fill=fillColor] (145.34, 26.29) rectangle (159.80, 40.75);
\end{scope}
\begin{scope}
\path[clip] (  0.00,  0.00) rectangle (252.94, 79.50);
\definecolor{fillColor}{RGB}{55,126,184}

\path[fill=fillColor] (146.05, 27.01) rectangle (159.09, 40.04);
\end{scope}
\begin{scope}
\path[clip] (  0.00,  0.00) rectangle (252.94, 79.50);
\definecolor{fillColor}{RGB}{255,255,255}

\path[fill=fillColor] (145.34,  9.59) rectangle (159.80, 25.29);
\end{scope}
\begin{scope}
\path[clip] (  0.00,  0.00) rectangle (252.94, 79.50);
\definecolor{fillColor}{RGB}{228,26,28}

\path[fill=fillColor] (146.05, 10.30) rectangle (159.09, 24.58);
\end{scope}
\begin{scope}
\path[clip] (  0.00,  0.00) rectangle (252.94, 79.50);
\definecolor{drawColor}{RGB}{0,0,0}

\node[text=drawColor,anchor=base west,inner sep=0pt, outer sep=0pt, scale=  0.80] at (165.30, 61.68) {\KaMinPar};
\end{scope}
\begin{scope}
\path[clip] (  0.00,  0.00) rectangle (252.94, 79.50);
\definecolor{drawColor}{RGB}{0,0,0}

\node[text=drawColor,anchor=base west,inner sep=0pt, outer sep=0pt, scale=  0.80] at (165.30, 46.22) {Two-Phase LP};
\end{scope}
\begin{scope}
\path[clip] (  0.00,  0.00) rectangle (252.94, 79.50);
\definecolor{drawColor}{RGB}{0,0,0}

\node[text=drawColor,anchor=base west,inner sep=0pt, outer sep=0pt, scale=  0.80] at (165.30, 30.77) {Graph Compression};
\end{scope}
\begin{scope}
\path[clip] (  0.00,  0.00) rectangle (252.94, 79.50);
\definecolor{drawColor}{RGB}{0,0,0}

\node[text=drawColor,anchor=base west,inner sep=0pt, outer sep=0pt, scale=  0.80] at (165.30, 19.01) {One-Pass Contraction};

\node[text=drawColor,anchor=base west,inner sep=0pt, outer sep=0pt, scale=  0.80] at (165.30, 10.37) {(\TeraPart)};
\end{scope}
\end{tikzpicture}
     \caption{Memory reduction when our optimizations are enabled one after another for \Instance{eu-2015} with $p = 96$ cores and $k = \numprint{30000}$
     blocks.
   }
   \label{fig:memory-burndown-plot}
 \end{figure}

Recently there has been substantial progress in scaling up graph partitioning systems by leveraging shared memory~\cite{KAMINPAR},\cite{MT-KAHYPAR},\cite{MT-METIS}, distributed~\cite{D-KAMINPAR, xtrapulp-journal} and GPU parallelism~\cite{JET}.
For a comprehensive overview of recent advances, we refer to a survey paper~\cite{more-recent-survey}.
Current solvers are 
sufficiently fast for many applications, but use a substantial amount of memory for auxiliary data structures to achieve this speed, in addition to the memory required to store the graph itself.
For example, partitioning the \Instance{eu-2015} web graph with 80.5 billion edges using the multilevel algorithm \KaMinPar~\cite{KAMINPAR} takes only 4 minutes on a 96-core machine, but uses 1.35\,TiB of RAM, see \KaMinPar in Figure~\ref{fig:memory-burndown-plot}.

 In pursuit of partitioning larger and larger graphs, we present and study optimizations that
 substantially reduce the memory footprint. Enabling the optimizations presented in this paper reduces the
 memory usage on \Instance{eu-2015} to around 0.1 TiB (see \TeraPart in \Cref{fig:memory-burndown-plot}) and speeds up the computation to just 80
 seconds. We further push the boundary of scale by partitioning synthetic graphs with one
 \textbf{trillion} undirected edges on a single machine in just under 8 minutes of computation,
 using around 900\,GiB of RAM. To the best of our knowledge, this is the first work in graph
 partitioning that successfully handles tera-scale graphs on a single machine, and the first work to
 employ the multilevel framework on tera-scale graphs. The only prior work reporting results with a trillion edges is \XtraPuLP~\cite{xtrapulp-journal}, which requires \numprint{8192} machines
 with 64\,GB each. Moreover, \XtraPuLP does not employ the multilevel framework, which results in
 substantially higher edge cuts: 5.56$\times$--\,68.44$\times$ in our experiments (confer
 \Cref{tbl:horeka}). Finally, we perform experiments with our techniques in the distributed
 version of \KaMinPar~\cite{D-KAMINPAR}, where we are able to scale to $2^{44}$ edges on 128 machines with 256\,GiB
 each.

On the technical side, we propose three novel optimizations to the components of the multilevel framework.
\begin{itemize}
\item We devise a \emph{two-phase label propagation} algorithm for the clustering step in the coarsening phase, which uses $O(n)$ memory instead of $O(np)$, where $n$ is the number of vertices and $p$ is the number of cores.
\item We propose a parallel contraction algorithm that constructs the representation of the contracted graph in \emph{one pass} over the input and eliminates storing the coarse graph a second time in a different representation.
\item For FM refinement, we propose a \emph{sparse gain table} that uses $O(m)$ memory instead of $O(nk)$.
\end{itemize}

Finally, we compress the input graph, which is an already established optimization in graph processing workloads~\cite{webgraph-framework, ligraplus}, though not yet considered for graph partitioning.
By integrating these optimizations into \KaMinPar, we obtain \TeraPart.

\section{Background}

 In this section, we provide additional context on the multilevel scheme and \KaMinPar, before
 determining the culprits of the high memory usage.

 \subsection{The Multilevel Scheme}

   Most state-of-the-art high-quality general-purpose solvers leverage the multilevel scheme~\cite{HENDRICKSON}. To \emph{coarsen}
   $G_0 \coloneqq G$, small vertex clusters are contracted repeatedly to obtain a hierarchy $G_0, G_1,
     G_2, \ldots$ of smaller but structurally similar graphs. After computing an initial partition on
   the smallest graph, the contractions are undone in reverse order, projecting the current partition
   to the next finer graph and improving the partition by moving vertices, using refinement
   algorithms. This approach is both faster and has higher solution quality than partitioning the
   input directly, as we perform global optimization on the top-level graph through local moves on the
   coarsened graphs.

 \subsection{KaMinPar}\label{sec:dkaminpar}

   We implement our optimizations in \KaMinPar~\cite{KAMINPAR, D-KAMINPAR}, a state-of-the-art
   multilevel graph partitioning system, which offers both shared-memory and distributed memory
   parallelism. For coarsening, it uses label propagation clustering, with additional two-hop
   matching~\cite{DBLP:conf/sc/LaSallePSSDK15} to ensure coarsening progress on irregular graphs.
   Starting from each vertex in its own cluster, label propagation visits vertices in random order in
   parallel; a vertex joins a cluster containing the plurality of its neighbors. For initial bipartitioning, it uses a portfolio of
   randomized sequential greedy graph growing heuristics 
   and $2$-way FM~\cite{FM}. In the refinement
   stage, \KaMinPar uses size-constrained label propagation~\cite{PARHIP}, starting with the given
   partition, and optionally shared-memory parallel localized $k$-way FM refinement~\cite{MT-KAHIP,
     MT-KAHYPAR}. 

   The distributed version of \KaMinPar splits the input graph into $p$ subgraphs, one for each process.
   Edges are assigned to the process responsible for the source vertex.
   If the target vertex of an edge is assigned to a different process, the vertex is replicated as a \emph{ghost} vertex (i.e., without outgoing edges).
   This requires some additional memory to store mappings between ghost vertices and their original vertices.
   Both coarsening and refinement are done using a distributed implementation of label propagation, which processes batches of vertices synchronously and in parallel.
   Balance violations are repaired in a subsequent rebalancing step.
   For initial partitioning, each process obtains a full copy of the coarsest graph and uses the shared-memory version of \KaMinPar for partitioning.

 \subsection{Memory Analysis}
   \begin{figure}[t]
     \input{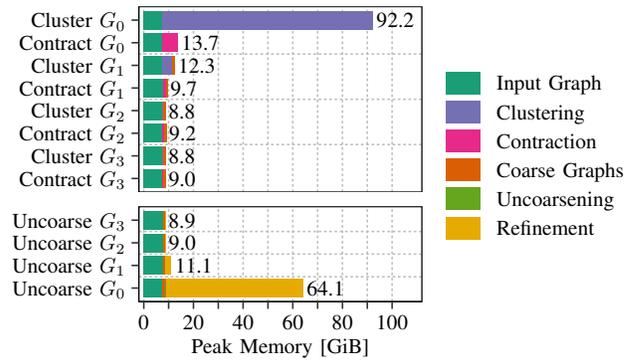}
     \caption{Memory consumption during the different phases of the \KaMinPar algorithm. Only the
       top level and three coarse levels are shown, as the memory consumption is barely reduced for
       the following levels. The measurements were carried out for \Instance{webbase2001} with $p = 96$
       cores and $k = 64$ blocks.
     }
     \label{fig:detailed-memory}
   \end{figure}

   In order to understand the areas for potential improvement, we analyze which component of \KaMinPar
   uses how much memory at which stage of the execution. Figure~\ref{fig:detailed-memory} breaks down
   the memory usage per stage and level using the \Instance{webbase2001}~\cite{NetworkRepository}
   graph as an example. For this analysis, we enabled the optional FM refinement.

   The top three memory peaks occur when working on the top-level graph $G_0$: 1) clustering in the
   coarsening stage, 2) FM refinement, 3) contraction. In the coarsening stage, \KaMinPar uses
   \numprint{92.2}\,GiB, \numprint{84.5} of which for auxiliary data structures in the clustering
   algorithm and \numprint{7.7} for the input graph. Two-phase label propagation reduces the auxiliary
   memory during clustering to \numprint{2.8}\,GiB. By compressing the input graph we only need
   \numprint{2.9}\,GiB to store it. Note that further memory savings from compressing coarser graphs
   (level $\geq 1$) are negligible, so that we only compress the input graph. During FM refinement,
   the auxiliary data structure uses \numprint{55.1}\,GiB of RAM, which we reduce to \numprint{5.6}\,GiB
   by employing our sparse gain table optimization. Note that label propagation refinement uses a
   negligible amount of memory, as it is proportional to $k$, rather than $n$. 
   Finally, one-pass
   contraction reduces the auxiliary memory from \numprint{6.0}\,GiB to \numprint{1.4}\,GiB. Combining the optimizations reduces
   the peak memory usage on \Instance{webbase2001} to \numprint{5.7}\,GiB when using label propagation
   refinement and to \numprint{9.5}\,GiB when using optional FM refinement.

   Given this analysis, a natural question is whether \KaMinPar is particularly wasteful and whether
   other partitioning algorithms, such as \MtMetis~\cite{MT-METIS, MT-METIS-REFINEMENT} are better. We
   found that \MtMetis uses 2$\times$--\,4$\times$ more memory than \KaMinPar, confer Figure~\ref{fig:ufm-time-memory} (middle).

\section{Graph Representation}

 The graphs are stored in the compressed sparse row (CSR) format. In the CSR format, the edges are
 stored in one contiguous array $\mathcal{E}$ of size $2m$. Additionally, an array $\mathcal{P}$ of
 size $n+1$ stores the beginning of each neighborhood, i.e., $\mathcal{E}[\mathcal{P}[u] :
     \mathcal{P}[u+1]]$ stores the neighbors of vertex $u$.

 To reduce the memory footprint of the input graph, we use a compressed representation instead of
 the CSR format. We further want to reduce the running time and memory space required to compress
 the input graph, which is why we compress the input graph in parallel in a single I/O-pass.

 \subsection{Compression Scheme}\label{ssec:compression}

   We employ compression techniques that enable fast decoding of each vertex's neighborhood. This is
   inspired by the success of \textsf{Ligra+}\cite{ligraplus} for graph processing workloads (which do
   not include graph partitioning). \textsf{Ligra+} uses gap encoding~\cite{webgraph-framework}
   combined with variable-length integer encoding (VarInt). With gap encoding, one sorts each
   neighborhood by increasing ID and stores only the difference to the previous neighbor ID. Since the
   gaps are often small, they can be represented in less than 8 bytes. VarInt is a variable-length
   byte codec, where 7 bits in a byte are used to represent part of a number and one bit (the
   continuation bit) indicates whether the 7 bits of the following byte belong to the current or a new
   number. Additionally, we combine this with interval encoding~\cite{webgraph-framework}, which lets
   us achieve less than one byte per compressed edge on some graphs, in contrast to \textsf{Ligra+}.
   With interval encoding, one stores consecutive intervals $\{x, x+1, x+2, \dots, x+\ell-1\}$ with $\ell \geq 3$ as a tuple
   $(x, \ell)$ rather than $\ell$ bytes representing gap $1$ each with gap encoding. This is particularly
   impactful in graphs with high locality in the neighbor IDs. For edge weights, we only employ 
   gap encoding combined with the VarInt codec because interval encoding is unlikely to yield better
   compression. 
   Since edge weights are not sorted, we store an additional sign bit.
   To enable parallel iteration over the neighborhood of a high-degree vertex (degree larger than \numprint{10000}), we split
   its neighbors into chunks of fixed length (size \numprint{1000}) that are encoded and decoded independently~\cite{ligraplus}. 

   Similar to the CSR format, the compressed neighborhoods are concatenated into one array storing the
   edges, and for each vertex we store a pointer to the beginning of its neighborhood. We store the
   edge weights of a neighborhood interleaved with its gaps and intervals. Moreover, due to
   compression, vertex degrees cannot be deduced from the range begin pointers. Hence, we store the
   first edge ID of a neighborhood as a VarInt at the start of the neighborhood to deduce the degrees.
   By storing the first edge ID instead of the degree, we can obtain the ID of each edge in a
   neighborhood during iteration, which is required by some parts of the \KaMinPar code.

 \subsection{Parallel Compression and Single-Pass I/O }\label{sec:io}

   The size of the uncompressed graph may exceed the available memory, so that loading the
   uncompressed graph and then compressing it from memory is not an option. Rather we want to perform the
   compression during I/O.

   One issue with CSR storage is that the size of the edge array must be known beforehand to allocate
   to the correct size. However, the size of the compressed edges is only known after loading and
   compressing them. The naive solution is to load the graph twice, computing the compressed size in
   the first pass, and storing the compressed graph in the second pass. However, often the graph I/O
   takes a similar time as partitioning itself, so ideally we want to load the graph only once.
   Growing arrays such as \texttt{std::vector} are not suitable either, since they require twice the
   memory to copy the data when growing.

   To avoid having to know the correct amount of memory in advance, we overcommit memory
   \cite{DBLP:conf/caip/WassenbergMS09},
   that is, we compute an upper bound on the memory consumption of the
   compressed edge array and request that amount. Moreover, we only touch the memory that we
   actually use. This technique works because the operating system only assigns virtual pages to
   physical pages when the corresponding page is touched. Thus, only the memory of the compressed edge
   array plus at most one page is physically backed by memory.

   To further speed up the I/O, we compress neighborhoods in parallel. However, to store a compressed neighborhood in the edge array, we have to know the size
   of the previous compressed neighborhoods, so that it can be written to the correct position. This
   is challenging, as this is essentially a prefix sum problem, which requires scanning the input twice when performed in parallel, and we only want to compress the
   neighborhoods once.
   We overcome this by using the two pass-approach locally. The threads work on packets of consecutive vertices that contain a similar number of edges and first compress them into a thread-local buffer. A thread that finished a packet waits until all preceding packets have
     found their memory requirement and updated the current edge array position accordingly. Then it increases the end position by the size of its buffer, marks the packet as finished, and proceeds to copy its buffer into the memory range preceding that position.

\section{Graph Coarsening}

In the coarsening stage, we first compute a disjoint clustering of the current level's graph using label propagation~\cite{LABEL_PROPAGATION} and then contract this clustering by merging all vertices in the same cluster.
Our goal is to reduce the amount of auxiliary memory needed by these algorithms during their execution.

\begin{algorithm2e}[t]
	\algsetup{linenosize=\tiny}
	\small
	\caption{Original Label Propagation Round}\label{algo:original-lp}
	\ParallelFor(){$u \in V$}{
		$\mathcal{R} \leftarrow $ rating map \;
		\For{$v \in N(u)$}{
			$\mathcal{R}[\mathcal{C}[v]] \leftarrow \mathcal{R}[\mathcal{C}[v]] + \omega(uv)$\;
		}
		$\mathcal{C}[u] \gets \displaystyle{\argmax_{c \in \mathcal{R}.\FuncSty{keys()}}} \mathcal{R}[c]$ \;
	}
\end{algorithm2e}

 \subsection{Label Propagation Clustering}\label{sec:two-phase-lp}

   Label propagation~\cite{LABEL_PROPAGATION} is an iterative clustering algorithm, where we move vertices into a neighboring cluster with the strongest connection in terms of edge weight.
   We represent the clustering as an array $\mathcal{C}$ of size $n$, mapping vertices to their assigned cluster. 
   Initially, each vertex starts in its own cluster, i.e., $\forall u \in V: \mathcal{C}[u] = u$.
   Then, we iterate through the vertices in parallel, determine the best cluster $c$ to join for a given vertex $u$, and update the clustering $C[u] \gets c$.
   See Algorithm~\ref{algo:original-lp} for pseudocode of one round of label propagation.   
   Each vertex is considered once per round.
   We perform five rounds before contracting the clustering.
  
   \subsubsection{Rating Maps}
  
   For a given vertex $u$, each cluster has an associated \emph{rating}: the sum of edge weights from $u$ to the cluster.
   To determine the best cluster to join, we iterate through $N(u)$ and aggregate the ratings in a \emph{rating map} data structure.
   The rating map can be implemented as either a hash table mapping cluster IDs to ratings or as a \emph{sparse array}, i.e., an array $\mathcal{A}$ of size $n$ storing the rating for cluster $c$ at position $c$ and a vector $L$ storing non-zero entries in $\mathcal{A}$, which is used to reset $\mathcal{A}$ after a vertex is processed.
   Hash tables use less memory (proportional to the maximum degree) but tend to have slower lookups in general, which is why sparse arrays are often preferred.
   Each thread needs its own rating map since we parallelize over the vertices.
   Hence, the rating maps are the single largest contributor to the peak memory footprint of clustering in \Cref{fig:detailed-memory}. 
   However, for vertices with small degree, the repeated local probing in hash tables can be faster due to better cache efficiency than random accesses to the sparse array~\cite{KAMINPAR, MT-KAHYPAR}.
   More precisely, the hash table size is proportional to the number of unique clusters in the neighborhood $\nc(u) \coloneqq |\{ \mathcal{C}[v] \mid v \in N(u) \}|$. 
   The hash table remains efficient as long as this number is small.
   
   \subsubsection{Two-Phase Label Propagation} 
   
   \begin{algorithm2e}[t]
    \algsetup{linenosize=\tiny}
    \small

    \SetKwInOut{Input}{Input}
    \SetKwInOut{Output}{Output}
       \caption{Two-Phase Label Propagation Round}\label{algo:2ps-lp}

    \SetKwFunction{FlushRatingMap}{FlushRatingMap}
	
	    \ParallelFor(\tcp*[f]{First Phase}){$u \in V$}{
	      $\mathcal{R} \leftarrow $ empty hash table [thread-local, fixed-capacity]\;
	
	      \For{$v \in N(u)$}{
	        $\mathcal{R}[\mathcal{C}[v]] \leftarrow \mathcal{R}[\mathcal{C}[v]] + \omega(uv)$\;
	        \BlankLine
	
	        \If{$|\mathcal{R}| \geq T_\text{bump}$}{
	          Bump $u$ and continue with next vertex
	        }
	      }
	      $\mathcal{C}[u] \gets \displaystyle{\argmax_{c \in \mathcal{R}.\FuncSty{keys()}}} \mathcal{R}[c]$ \;
	    }
	
	    $\mathcal{A} \leftarrow $ allocate zero-initialized array of size $n$\;
	    \For(\tcp*[f]{Second Phase}){each bumped vertex $u$}{
	      \ParallelFor{$v \in N(u)$}{
	        $\mathcal{R}_t[\mathcal{C}[v]] \leftarrow \mathcal{R}_t[\mathcal{C}[v]] + \omega(uv)$\;
	        \BlankLine
	
	        \If{$|\mathcal{R}_t| \geq T_\text{bump}$}{
	          \FlushRatingMap{$\mathcal{A}$, $\mathcal{R}_t$, $L_t$}\;
	        }
	      }
	      $\FlushRatingMap{$\mathcal{A}$, $\mathcal{R}_t$, $L_t$} \; \forall t \in [p]$\;
	      \BlankLine
	
	      $\mathcal{C}[u] \gets \displaystyle{\argmax_{c \in \bigcup_{t \in [p]} L_t}} \mathcal{A}[c]$ \;
	      $\mathcal{A}[c] \gets 0 \; \forall c \in \bigcup_{t \in [p]} L_t$ \;
	    }

	\Fn{\FlushRatingMap{$\mathcal{A}$, $\mathcal{R}_t$, $L_t$}} {
		\For{$(c, w) \in \mathcal{R}_t$}{
			$w_\text{prev} \leftarrow \mathcal{A}[c] \underset{\text{atomic}}{\pluseq} w$\; \label{algo:2ps-lp:atomic}
			\If{$w_\text{prev} = 0$}{ \label{algo:2ps-lp:track}
				Append $c$ to thread-local vector $L_t$\;
			}
		}
		$\mathcal{R}_t.\FuncSty{clear()}$\;
	}
  \end{algorithm2e}
   
   This motivates a simple adaptation to substantially reduce memory and preserve performance.
   We split the label propagation round into two phases as shown in Algorithm~\ref{algo:2ps-lp}.
   First we try to process all vertices in parallel with small fixed-capacity hash tables (no dynamic growth). 
   While processing a vertex $u$, if $\nc(u)$ exceeds a threshold $T_\text{bump}$, we stop and \emph{bump} it to the second phase.
   In the second phase, we process the few bumped vertices sequentially with a sparse array as the rating map and employ parallelism over the edges.
   As we only use one sparse array, we need just $O(n + p \cdot T_\text{bump})$ instead of $O(n \cdot p)$ memory, where $p$ is the number of processors.
   The term $p \cdot T_\text{bump}$ is negligible compared to $n$ on current machines and graphs where reducing memory usage matters.
   Since the degree of a bumped vertex is at least $T_\text{bump}$, there is sufficient work to justify parallelism over the edges.

  \subsubsection{Parallel Sparse Array}
  
  With parallelism over edges there is a race condition on the cluster ratings in the sparse array $\mathcal{A}$.
  Therefore, we use atomic fetch-add instructions to aggregate the scores safely, see line~\ref{algo:2ps-lp:atomic} of Algorithm~\ref{algo:2ps-lp}.
  We implement the list $L$ of non-zero entries as thread-local buffers $L_t$.
  To prevent duplicate cluster IDs in $L = \bigcup_{t \in [p]} L_t$, only the thread that raises the rating of cluster $c$ from $\mathcal{A}[c] = 0$ to $\mathcal{A}[c] > 0$ tracks $c$ in its buffer.
  The atomic fetch-add instruction returns the value just \emph{before} the operation, which we check in line~\ref{algo:2ps-lp:track}.
  
  Another concern is contention.
  We only process vertices with high $\nc(u)$ in the second phase, so that rating contributions are spread across many clusters and thus memory locations, which helps to some extent.
  However, it is still possible that few clusters receive a majority of the atomic increments.
  To reduce contention, we leverage the hash tables from the first phase as intermediate buffers, to perform fewer atomic updates overall.
  Once a hash table reaches capacity or all neighbors are traversed, we flush the hash table by applying its entries to $\mathcal{A}$ with the atomic fetch-add instruction.

 \subsection{Contraction}\label{ssec:contraction}

   \begin{figure}
   	\includegraphics[width=\linewidth]{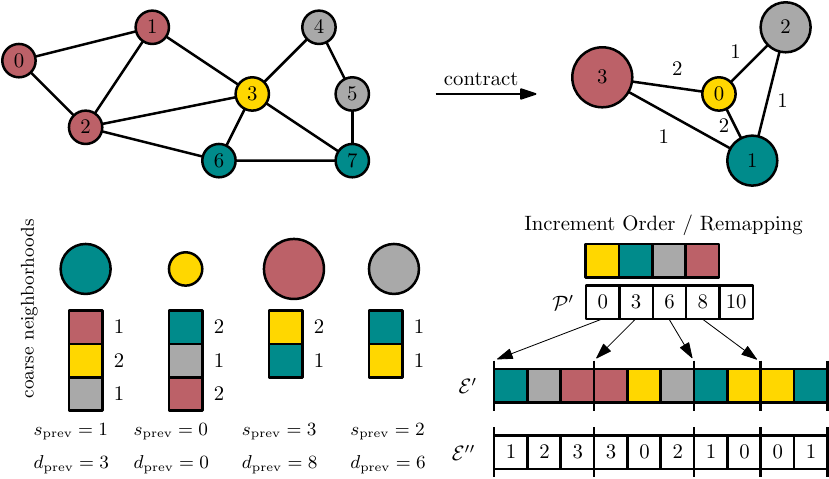}
   	\caption{Illustration of the contraction step. After computing a coarse neighborhood, we increment the counters $s$ and $d$ to obtain the start position $d_\text{prev}$ in $\mathcal{E}'$ and new coarse vertex ID $s_\text{prev}$. Once all neighborhoods are inserted, we remap the old cluster IDs (denoted as colors) to the new coarse vertex IDs (numbers).}\label{fig:contraction}
   \end{figure}
   
   Given the clustering $\mathcal{C}$, we want to construct the CSR representation $(\mathcal{P}', \mathcal{E}')$ of the contracted graph $G'= (V', E')$ to obtain the next level in the multilevel hierarchy.
   Let $\mathcal{C}^{-1}[c]= \{ u \in V \mid \mathcal{C}[u] = c \}$ be the set of vertices in the cluster with ID $c$.
   Recall that the clusters in $G$ correspond to vertices in $G'$ and intra-cluster edges $(u,v) \in E$ with $\mathcal{C}[u] \neq \mathcal{C}[v]$ in $G$ become edges $(\mathcal{C}[u], \mathcal{C}[v])$ in $G'$.
   The weight of an edge $(a,b)$ in $G'$ is $\omega'(a,b) = \sum_{u \in \mathcal{C}^{-1}[a], v \in \mathcal{C}^{-1}[b], (u,v) \in E} \omega(u,v)$, i.e., the sum of weights of duplicate edges after remapping $(u,v)$ to $(\mathcal{C}[u], \mathcal{C}[v])$.
   For each cluster $a$, we compute its outgoing edges in $G'$ by iterating over the vertices $u \in \mathcal{C}^{-1}[a]$ and scanning all outgoing edges $(u,v)$ of $u$ in $G$, adding $\omega(u,v)$ to the edge weight $\omega'(a, \mathcal{C}[v])$.
   Note the similarity to rating aggregation in label propagation.
      
   \subsubsection{Two-Phase Aggregation}
   
   In \Cref{fig:detailed-memory}, we observe that after label propagation coarsening and FM refinement, contraction is the third largest contributor to the peak memory.
   While sparse arrays only require $n' = |V'| \ll n$ number of entries in the contraction step, this is a good avenue for optimization.
   Therefore, we also apply our two-phase aggregation approach from Section~\ref{sec:two-phase-lp} here.
   
   In the first phase, we employ parallelism over the clusters.
   Within a cluster $c$, we scan through its contained vertices $u \in \mathcal{C}^{-1}[c]$ and their neighborhoods sequentially.
   For each neighbor $v \in N(u)$, we add $\omega(u,v)$ to the edge weight $\omega'(c, C[v])$.
   In the second phase, we process one cluster at a time, but employ parallelism over $\mathcal{C}^{-1}[c]$, and potentially the neighborhoods if the degree in $G$ is sufficiently high.
   
   \subsubsection{One-Pass Contraction}\label{sec:direct-csr}
   
   The second piece of auxiliary memory used in contraction is a set of temporary buffers storing $E'$ during aggregation; before the edges are copied to $\mathcal{E}'$.
   The difficulty with using CSR in the parallel setting is the necessity to know each vertex degree and offset $\mathcal{P}'$ (prefix sum over the degrees), before edges can be written to $\mathcal{E}'$ in parallel.
   In the sequential setting, one can simply compute the prefix sum at the same time as appending edges.
   In the parallel setting however, the degrees of vertices whose edges should appear earlier in $\mathcal{E}'$ may not be known yet.
   Once all coarse edges are computed, we can compute the prefix sum over degrees to obtain the offsets $\mathcal{P}'$, and then copy the edges to the appropriate locations in $\mathcal{E}'$.
   In the remainder of this section, we describe how to avoid the buffers, so that we store the coarse graph only once, in CSR format, and only compute the coarse edges once\footnote{The simplest approach is to compute the coarse edges twice: once to count the degrees, and a second time to place the edges in $\mathcal{E}'$.
   	As this doubles the running time, this is not an option.}.
         
   Similar to the single-pass I/O in \Cref{sec:io}, we employ virtual memory overcommitment to allocate $2m$ entries for $\mathcal{E}'$ without physical memory backing, as we do not know the true $m' = |E'|$ yet.
   We append newly computed coarse edges to $\mathcal{E}'$, thus only $2m'$ entries plus at most one page of memory are physically allocated.
   Note that we use $2m'$, as we compute and store each edge in both directions.
   
   To some extent, our approach to resolve the prefix sum issue mimics the sequential approach, but uses atomic instructions to fix race conditions and uses buffering to hide contention from the use of atomics.
   Let $d \gets 0$ be an index counting the number of edges already inserted  to $\mathcal{E}'$ and $s \gets 0$ denote the number of coarse vertices already processed.
   At the end we will have $d = 2m'$ and $s = n$.
   See also \Cref{fig:contraction} for an illustration.
   
   The following applies to the first phase, where coarse vertices are processed in parallel.
   For each coarse vertex $u' \in V'$, we start computing its coarse edges and store them in a thread-local hash table $R_t$. As soon as $|R_t| \geq T_\text{bump}$, we bump $u'$ to the second phase.
   Otherwise, $R_t$ contains all edges of $u'$.
   We increment $d \underset{\text{atomic}}{\pluseq} |R_t|$ and capture the value $d_\text{prev}$ immediately before the increment.
   Then we copy $R_t$ to the range $\mathcal{E}'[d_\text{prev}: d_\text{prev} + |R_t|]$ (see bottom of \Cref{fig:contraction}).
   
   Note that due to parallelism, the neighborhoods in $\mathcal{E}'$ are out of order, i.e., are not stored in the same order as the coarse vertex IDs.
   Thus, one can either store a begin and end pointer to the neighborhood of each vertex (which requires twice the memory), or relabel the vertices, so that the neighborhoods of consecutive vertex IDs are consecutive in $\mathcal{E'}$.
   To this end, we increment $s  \underset{\text{atomic}}{\pluseq} 1$ for each coarse vertex processed; in a transaction combined with the update $d \underset{\text{atomic}}{\pluseq} |R_t|$, as they must be consistent.
   Again, we capture $s_\text{prev}$ as the value before the transaction and save the beginning of the neighborhood in $\mathcal{P}'[s_\text{prev}] \gets d_\text{prev}$.
   Additionally, we store the new vertex ID $s_\text{prev}$ so that we can remap the endpoints of the coarse edges at the end, thus avoiding to shuffle the neighborhoods in $\mathcal{E}'$.
   
   To synchronously update $d$ and $s$ in a transaction, we employ the double-width compare-and-swap instruction~\cite{INTEL-INSTRUCTION-SET-REFERENCE}.
   The two counters are stored as a 128-bit integer, with $d$ stored in the lower 64 bits and $s$ in the upper bits.
   We implement the transaction as a compare-and-swap loop, where we extract, update and repack the values in the loop body.
   
   In order to perform fewer compare-and-swap instructions and thus reduce contention, we increment the dual counter for several coarse vertices at once. To implement this, we store the neighbors of multiple coarse vertices in another fixed-capacity buffer $B_t$. Once the buffer reaches capacity, or no coarse vertices are left to process, we increment $d$ by $|B_t|$ and $s$ by the number of coarse neighborhoods stored in the buffer.
   In the second phase where coarse high-degree vertices are processed sequentially, there is no need to atomically update $d$ and $s$.

\section{Space-Efficient Gain Tables}\label{sec:fm}

A common performance optimization in refinement algorithms is a \emph{gain table}.
For each pair of a vertex $u \in V$ and a block $V_i$, we cache the value $\omega(u, V_i) = \sum_{(u,v) \in E, v \in V_i} \omega(u,v)$ in the table, which we call the \emph{affinity} of a vertex to a block.
Let $\Partition(u)$ denote the block to which $u$ is assigned.
With cached affinity values, the gain of moving $u$ to $V_i$ can be computed as $\omega(u, V_i) - \omega(u, \Partition(u))$.
After moving a vertex $u$ from block $V_s$ to $V_t$, we update the affinity for each neighbor $v$ of $u$ as $\omega(v, V_s) \minuseq \omega(u,v)$ and $\omega(v, V_t) \pluseq \omega(u,v)$ using atomic fetch-add instructions.
This optimization is standard practice for algorithms such as FM~\cite{FM} where vertices' gains will be inspected substantially more often than moves performed, such that the overhead of updating the gain table after a move is lower than computing the gain from scratch, each time the vertex is inspected.

Unfortunately, the standard implementation of gain tables uses $k$ entries per vertex for a total of $O(nk)$ memory, which is infeasible for very large graphs and even moderate values of $k$.
Instead, we would like to use $O(m)$ memory.
The idea is to use the standard implementation with $k$ entries only for vertices with $\deg(v) > k$. 
For vertices with lower degree, we use tiny linear-probing hash tables with fixed capacity $\Theta(\deg(v))$, as the vertex can be adjacent to at most $\deg(v)$ different blocks. The affinity value for non-adjacent blocks is zero, which we do not store explicitly.
This leads to a memory footprint of $O(\sum_{v \in V} \min (\deg(v), k)) \subset O(m)$.

The updates for affinity values of low-degree vertices cannot be applied with atomic fetch-add instructions, as values that drop to zero must be removed.
Since we anticipate far more queries than deletions, we move up elements to close gaps in the probing order upon deletion~\cite{SMDD19}. 
Therefore, positions in the hash table are not stable such that each hash table must be protected by a spinlock.
Somewhat surprisingly, these overheads have a minor impact on running time, as we demonstrate in the experiments.

The memory for all affinity values is allocated in one contiguous array.
Note that the affinity values of a vertex are upper bounded by its total incident edge weight $U$.
To reduce the memory footprint of the gain table further, we choose a variable width $w$ (8, 16, 32 or 64 bits) for the entries of each vertex as the smallest value $w > \log_2(U)$. 
Furthermore, for each vertex, we keep a pointer to the beginning of the slice of memory storing its affinity values.

\section{Experiments}
We have integrated the described algorithms into \KaMinPar~\cite{KAMINPAR}. 
Additionally, we have integrated the graph compression techniques into its distributed-memory counterpart \dKaMinPar~\cite{D-KAMINPAR}, which uses message passing for parallelization. 
Thus, our other optimizations are not applicable here.
We denote the resulting shared-memory resp. distributed-memory algorithms as \TeraPart resp. \xTeraPart. 
We compile all codes using \gpp-13.2.0 with flags \texttt{-mtune=native -march=native -msse4.1 -mcx16} and use Intel TBB~\cite{TBB} resp. Open MPI 4.0~\cite{OpenMPI} for parallelization.

\begin{figure*}[t]
    \centering
    \input{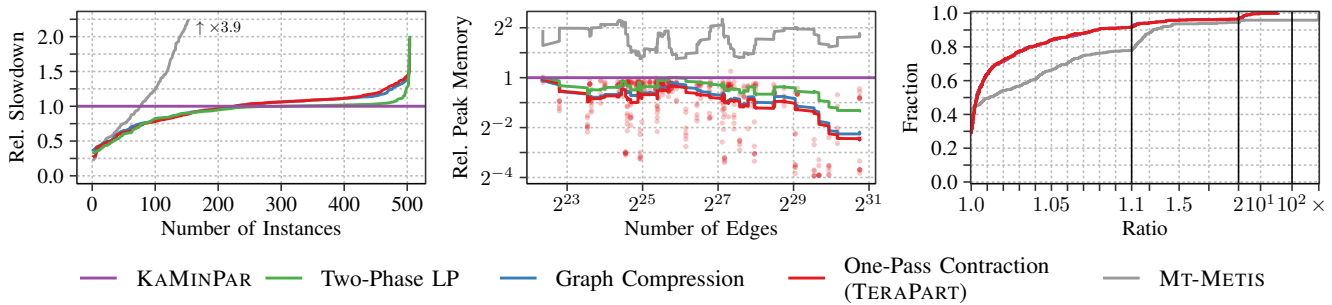}
\begin{tikzpicture}[x=1pt,y=1pt]
\definecolor{fillColor}{RGB}{255,255,255}
\begin{scope}
\definecolor{fillColor}{RGB}{255,255,255}

\path[fill=fillColor] ( 25.49,238.44) rectangle (480.40,267.45);
\end{scope}
\begin{scope}
\definecolor{fillColor}{RGB}{255,255,255}

\path[fill=fillColor] ( 30.99,243.94) rectangle ( 48.33,261.95);
\end{scope}
\begin{scope}
\definecolor{drawColor}{RGB}{152,78,163}

\path[draw=drawColor,line width= 1.1pt,line join=round] ( 32.72,252.94) -- ( 46.60,252.94);
\end{scope}
\begin{scope}
\definecolor{fillColor}{RGB}{255,255,255}

\path[fill=fillColor] (101.14,243.94) rectangle (118.49,261.95);
\end{scope}
\begin{scope}
\definecolor{drawColor}{RGB}{77,175,74}

\path[draw=drawColor,line width= 1.1pt,line join=round] (102.88,252.94) -- (116.75,252.94);
\end{scope}
\begin{scope}
\definecolor{fillColor}{RGB}{255,255,255}

\path[fill=fillColor] (189.82,243.94) rectangle (207.17,261.95);
\end{scope}
\begin{scope}
\definecolor{drawColor}{RGB}{55,126,184}

\path[draw=drawColor,line width= 1.1pt,line join=round] (191.56,252.94) -- (205.43,252.94);
\end{scope}
\begin{scope}
\definecolor{fillColor}{RGB}{255,255,255}

\path[fill=fillColor] (298.90,243.94) rectangle (316.25,261.95);
\end{scope}
\begin{scope}
\definecolor{drawColor}{RGB}{228,26,28}

\path[draw=drawColor,line width= 1.1pt,line join=round] (300.64,252.94) -- (314.51,252.94);
\end{scope}
\begin{scope}
\definecolor{fillColor}{RGB}{255,255,255}

\path[fill=fillColor] (417.93,243.94) rectangle (435.27,261.95);
\end{scope}
\begin{scope}
\definecolor{drawColor}{gray}{0.60}

\path[draw=drawColor,line width= 1.1pt,line join=round] (419.66,252.94) -- (433.54,252.94);
\end{scope}
\begin{scope}
\definecolor{drawColor}{RGB}{0,0,0}

\node[text=drawColor,anchor=base west,inner sep=0pt, outer sep=0pt, scale=  0.88] at ( 53.83,249.91) {\KaMinPar};
\end{scope}
\begin{scope}
\definecolor{drawColor}{RGB}{0,0,0}

\node[text=drawColor,anchor=base west,inner sep=0pt, outer sep=0pt, scale=  0.88] at (123.99,249.91) {Two-Phase LP};
\end{scope}
\begin{scope}
\definecolor{drawColor}{RGB}{0,0,0}

\node[text=drawColor,anchor=base west,inner sep=0pt, outer sep=0pt, scale=  0.88] at (212.67,249.91) {Graph Compression};
\end{scope}
\begin{scope}
\definecolor{drawColor}{RGB}{0,0,0}

\node[text=drawColor,anchor=base west,inner sep=0pt, outer sep=0pt, scale=  0.88] at (321.75,254.67) {One-Pass Contraction};

\node[text=drawColor,anchor=base west,inner sep=0pt, outer sep=0pt, scale=  0.88] at (321.75,245.16) {(\TeraPart)};
\end{scope}
\begin{scope}
\definecolor{drawColor}{RGB}{0,0,0}

\node[text=drawColor,anchor=base west,inner sep=0pt, outer sep=0pt, scale=  0.88] at (440.77,249.91) {\MtMetis};
\end{scope}
\end{tikzpicture}
    \caption{
        Relative running times (left) and peak memory (middle) on Benchmark Set A for \TeraPart relative to \KaMinPar when enabling the following optimizations one after the other: (i) two-phase label propagation, (ii) graph compression, and (iii) one-pass cluster contraction.
        For peak memory, we plot the per-instance ratios for \TeraPart with a right-aligned rolling geometric mean over 50 instances (shown for all algorithms).
        We include \MtMetis as a reference point, which is $3.9\times$ slower than \KaMinPar on average (excluding instances for which it did not produce a result). 
        The performance profile (right) compares the solution quality of \KaMinPar, \TeraPart and \MtMetis.
        Note that the curves of \KaMinPar and \TeraPart lie on top of each other, indicating that our optimizations do not affect solution quality. 
    }
    \label{fig:ufm-time-memory}
\end{figure*}

\paragraph*{Setup} 
We perform most of our experiments on a machine equipped with a 96-core AMD EPYC 9684X processor clocked at 2.55\,GHz and \numprint{1536}\,GiB of main memory, running Ubuntu 24.04. 
Unless stated otherwise, we obtain our results using all 96 cores. 
Additionally, we evaluate \xTeraPart on the \HoreKa high-performance cluster, where each node is equipped with two 39-core Intel Xeon Platinum 8368 processors and 256\,GiB of main memory. 
The compute nodes are connected by an InfiniBand 4X HDR 200\,Gbit/s network with approx. 1\,$\mu$s latency. 
We only use 64 out of the available 78 cores per node since the \Partitioner{KaGen}~\cite{KAGEN} graph generator requires the number of cores to be a power of two.

\paragraph*{Instances} 
To cover a diverse range of medium sized graphs, we use the benchmark instances of Ref.~\cite{unconstrained-fm} for our shared-memory experiments (Set A). 
This benchmark set consists of \RnumGraphsInSetA{} graphs obtained from the SuiteSparse Matrix Collection~\cite{SPM}, Network Repository~\cite{NetworkRepository}, graphs created by compressing texts from the Pizza\&Chill corpus~\cite{PIZZA-CHILI-TEXTS, TEXT-RECOMPRESSION} as well as synthetic graphs~\cite{KAGEN}. 
Their size ranges from \RsmallestNumEdgesInSetAMillion{} million to \RlargestNumEdgesInSetABillion{} billion undirected edges.
The graphs are mostly unweighted except for the six graphs from the text compression class. 
More details are available in Ref.~\cite{unconstrained-fm} and shown in Appendix~\Cref{fig:benchmark-set-a}.
Additionally, we perform experiments on several huge web graphs (see \Cref{table:hwg-stats}) from the Laboratory for Web Algorithmics~\cite{webgraph-framework, BRSLLP, BMSB}, namely \Instance{gsh-2012}, \Instance{clueweb12}, \Instance{uk-2014} and \Instance{eu-2015}, as well as the \Instance{hyperlink}(-2012)~\cite{HYPERLINK-2012} graph (Set B).
We converted these graphs to undirected graphs by adding missing reverse edges and removing any self-loops.

We further use families of random geometric (\Instance{rgg2D}) and hyperbolic (\Instance{rhg}) graphs generated using the \textsc{KaGen}~\cite{KAGEN} graph generator. 
\Instance{rgg2D} graphs do not have any high degree vertices and resemble mesh-like graphs, whereas \Instance{rhg} graphs have a skewed power-law degree distribution and model real-world social networks.
We generate these graphs with average degree $d = 256$ and power-law exponent $\gamma = 3.0$ (\Instance{rhg} only).

\begin{table}[t]
    \centering
    \caption{
        Benchmark Set B: number of vertices \(n\), number of undirected edges \(m\), average degree \(d(G)\) and max degree \(\Delta(G)\).
    }
    \begin{tabular}{l|rrrr}
        Graph $G$ & $n$                   & $m$                     & $d(G)$    & $\Delta(G)$         \\
        \midrule
        \Instance{gsh-2015}  & \numprint{988490691}  & \numprint{25690705118}  & \numprint{51}  & \numprint{58860305} \\
        \Instance{clueweb12} & \numprint{978408098}  & \numprint{37372179311}  & \numprint{76}  & \numprint{75611696} \\
        \Instance{uk-2014}   & \numprint{787801471}  & \numprint{42464215550}  & \numprint{107} & \numprint{8605492}  \\
        \Instance{eu-2015}   & \numprint{1070557254} & \numprint{80528515647}  & \numprint{150} & \numprint{20252259} \\
        \Instance{hyperlink} & \numprint{3563602789} & \numprint{112243992243} & \numprint{71}  & \numprint{45676832} \\
    \end{tabular}
    \label{table:hwg-stats}
\end{table}

\paragraph*{Methology} 
We consider the combination of a graph and the number of blocks $k$ as an \emph{instance}. 
Unless stated otherwise, we use $\varepsilon = 3\%$, $k \in \{8, 37, 64, 91, 128, \numprint{1000}, \numprint{30000}\}$ and perform $5$ repetitions for each instance using different seeds. 
We aggregate running times, peak memory consumption and edge cuts for the same instance using the arithmetic mean over all seeds. 
When aggregating over multiple instances, we use the harmonic mean for relative speedups and the geometric mean for all other metrics.

Except for the generated graphs used in \Cref{sec:exp:horeka}, all graphs are stored on disk in an uncompressed binary format. 
Compression happens on-the-fly during the streaming process into memory, introducing some overhead compared to directly loading the graphs into memory.
To keep this overhead low, we have parallelized the compression process, refer to \Cref{sec:io} for details.
For instance, on \Instance{eu-2015}, it takes \RioTimeCompressedSeqS{}\,s and \RioTimeUncompressedSeqS{}\,s to load the graph sequentially with and without on-the-fly graph compression from a RAID 0 with two Micron 7450 PRO NVMe SSDs.
With 96 cores, I/O times reduce to \RioTimeCompressedParS{}\,s and \RioTimeUncompressedParS{}\,s, respectively.
Since the overhead with parallel I/O is minimal, we generally exclude I/O times from the measurements.
Including I/O times would also be unfair to \MtMetis, which reads graphs in a text format and thus exhibits overheads due to parsing.

To compare the edge cuts of different algorithms, we use \emph{performance profiles}~\cite{PERFORMANCE-PROFILES}. 
Let $\mathcal{A}$ be the set of all algorithms we want to compare, $\mathcal{I}$ the set of instances, and $\cut_{A}(I)$ the edge cut of algorithm $A \in \mathcal{A}$ on instance $I \in \mathcal{I}$. 
For each algorithm $A$, we plot the fraction of instances $\frac{|\mathcal{I}_{A}(\tau)|}{|\mathcal{I}|}$ ($y$-axis) where 
$\mathcal{I}_{A}(\tau) \coloneqq \{ I \in \mathcal{I} \mid \cut_A(I) \leq \tau \cdot \min_{A' \in \mathcal{A}}\cut_{A'}(I) \}$ and $\tau$ is on the $x$-axis. 
Achieving higher fractions at lower $\tau$-values is considered better.
For $\tau = 1$, the $y$-value indicates the fraction of instances for which an algorithm performs best, while $\tau > 1$ shows the robustness of an algorithm. 
For example, we might prefer an algorithm that is only best in 49\% of the instances but never more than 1\% off the best solution while the other algorithm produces much worse partitions for 49\% of the inputs.

\subsection{Shared-Memory with Label Propagation Refinement}

\subsubsection{Medium-Sized Graphs}
In \Cref{fig:ufm-time-memory}, we evaluate the impact of the proposed optimizations on running time, peak memory and edge cut on Benchmark Set A.
We start with the \KaMinPar baseline and enable the optimizations one by one in the following order: 

\begin{enumerate}
    \item[(i)] two-phase label propagation (\Cref{sec:two-phase-lp}), 
    \item[(ii)] graph compression (\Cref{ssec:compression}) and 
    \item[(iii)] \TeraPart: one-pass cluster contraction (\Cref{ssec:contraction}).
\end{enumerate}

Two-phase label propagation improves both memory consumption and execution time of \KaMinPar, reducing its average memory peak by \RufmLpTpVsSpHeapPerc{}\% from \RufmLpSpHeapGiB{}\,GiB to \RufmLpTpHeapGiB{}\,GiB while improving its running time by \RufmLpTpVsSpTimePerc{}\% from \RufmLpSpTimeS{}\,s to \RufmLpTpTimeS{}\,s.
This speedup is not unexpected, since two-phase label propagation achieves better load balance through parallel processing of large neighborhoods. 
This is beneficial for graphs $G$ with high maximum degree (i.e., $\Delta(G) \ge T_{\text{bump}} = \numprint{10000}$), where we observe an improvement in running time by \RufmLpTpVsSpTimeLargeDegPerc{}\%.

Gap and interval encoding achieve an average compression factor of \RufmCompressionRatio{} across the entire benchmark set, although compression ratios vary greatly depending on the application domains of the graphs. 
For instance, compression ratios range from slightly below 1 for \Instance{kmer\_*} graphs~\cite{SPM} to \RufmCompressionRatioFiniteElement{} for graphs derived from finite element simulations.
We report per-graph compression ratios in Appendix~\Cref{fig:compression-ratios}.
In our partitioning experiment, these ratios translate to a \RufmLpGcVsTpHeapPerc{}\% reduction in peak memory, while increasing running time by just \RufmLpTpVsGcTimePerc{}\%.

Finally, we obtain \TeraPart by switching to our one-pass contraction algorithm, which further reduces both peak memory consumption (to \RufmLpUcHeapGiB{}\,GiB) and running time (to \RufmLpUcTimeS{}\,s) by \RufmLpUcVsGcHeapPerc{}\% and \RufmLpUcVsGcTimePerc{}\%, respectively.
With all three optimizations enabled, \TeraPart consumes \RufmLpTeraPartVsKaMinParHeapPerc{}\% less memory than \KaMinPar, while being \RufmLpTeraPartVsKaMinParTimePerc{}\% faster. 
Perhaps surprisingly, we can therefore conclude that \TeraPart does not only optimize the memory consumption of \KaMinPar, but also improves its running time.

As shown in \Cref{fig:ufm-time-memory} (middle), the memory savings of \TeraPart are more pronounced on larger graphs. 
Indeed, considering only graphs with $m \ge 10^8$ edges (\RufmLpNumLargeInstances{} out of \RufmLpNumInstances{} instances), average peak memory reduction increases to \RufmLpLargeTeraPartVsKaMinParHeapPerc{}\%, and further increases to \RufmLpVeryLargeTeraPartVsKaMinParHeapPerc{}\% when considering only graphs with $m \ge 10^9$ edges (\RufmLpNumVeryLargeInstances{} instances).

In the performance profile \Cref{fig:ufm-time-memory} (right), we compare the solution quality of \KaMinPar and \TeraPart.
We see that both curves lie on top of each other, i.e., our optimizations do not affect the solution quality of the partitioner (the average edge cuts of both partitioners are within \RufmLpKaMinParVsTeraPartCutPerc{}\% of each other).

To contrast our results against well established graph partitioners, we also include \MtMetis in \Cref{fig:ufm-time-memory}.
\MtMetis fails to produce any partition for the three largest graphs\footnote{On \Instance{twitter-2010}, \Instance{com-Friendster} and \Instance{sk-2005}.} in our medium-sized benchmark set for all values of $k$.
For the remaining graphs, we find that \MtMetis is on average \RufmLpMtMetisVsKaMinParTimeRatio{} times slower than \KaMinPar while consuming \RufmLpMtMetisVsKaMinParRSSRatio{} times more memory than even our unoptimized \KaMinPar baseline.
Compared to the optimized \TeraPart, \MtMetis is \RufmLpMtMetisVsTeraPartTimeRatio{} times slower while consuming \RufmLpMtMetisVsTeraPartRSSRatio{} times more memory on average.
We further observe that \MtMetis does not always respect the balance constraint, producing imbalanced partitions for \RufmLpMtMetisNumImbalancedInstances{} out of \RufmLpNumInstances{} instances.
However, looking at \Cref{fig:ufm-time-memory} (right), we see that even when ignoring the infeasibility of the partitions, \TeraPart still finds partitions of similar quality that are also balanced.

\begin{figure}[t]
    \centering
    \input{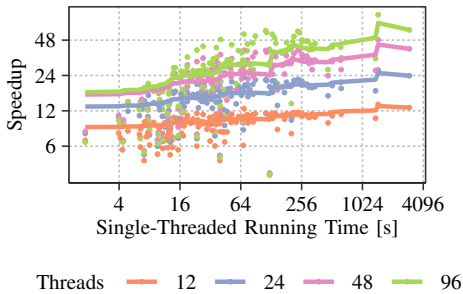}
    \caption{
        Self-relative speedups for \TeraPart with $p \in \{12, 24, 48, 96\}$ cores and $k \in \{64, \numprint{30000}\}$.
        We sort instances $i$ by their sequential running time $t^i$ and plot the cumulative harmonic mean speedup of all instances with sequential running time $t^i \ge t$ at position $t$.
    }
    \label{fig:ufm-scalability}
\end{figure}

\Cref{fig:ufm-scalability} shows the scalability of \TeraPart for 
$k \in \{64, \numprint{30000}\}$ 
on the medium-sized graphs of Benchmark Set A using $p \in \{12, 24, 48, 96\}$ cores of our shared-memory machine.
The overall harmonic mean speedup is \RufmLpSpeedupTwelve{} for $p = 12$, \RufmLpSpeedupTwentyFour{} for $p = 24$, \RufmLpSpeedupFortyEight{} for $p = 48$ and \RufmLpSpeedupNinetySix{} for $p = 96$.
The speedup achieved with $96$ cores over $48$ is somewhat limited when averaging over all instances.
This is mostly due to the initial partitioning phase, which can only make full use of the available parallelism once the graph is partitioned into a sufficiently large number of blocks $k' \ge p$.
On larger graphs which require at least 64\,s of sequential processing time (\RufmLpSpeedupNumMediumInstances{} out of \RufmLpSpeedupNumInstances{} instances), our speedups increase to \RufmLpSpeedupTwelveMedium{}, \RufmLpSpeedupTwentyFourMedium{}, \RufmLpSpeedupFortyEightMedium{} and \RufmLpSpeedupNinetySixMedium{} for $12$, $24$, $48$ and $96$ cores, respectively.
For the \RufmLpSpeedupNumLargeInstances{} instances which require at least 256\,s of sequential processing time, we measure a speedup of \RufmLpSpeedupNinetySixLarge{} on $p = 96$ cores.
Note that \TeraPart does not perform any expensive arithmetic operations and is limited by memory bandwidth.
Thus, perfect speedups are not possible.

\subsubsection{Huge Web Graphs}

\begin{figure*}[t]
    \centering
    \input{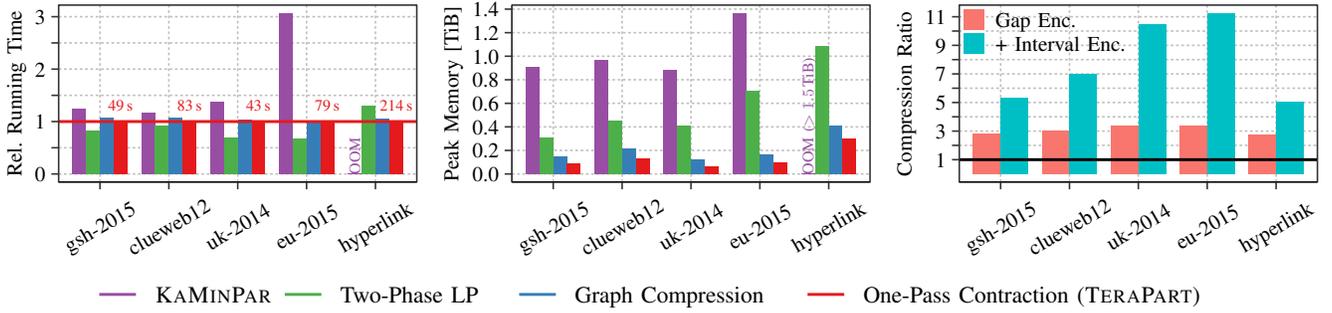}
\begin{tikzpicture}[x=1pt,y=1pt]
\definecolor{fillColor}{RGB}{255,255,255}
\begin{scope}
\definecolor{fillColor}{RGB}{255,255,255}

\path[fill=fillColor] ( 34.22,238.77) rectangle (471.67,267.12);
\end{scope}
\begin{scope}
\definecolor{fillColor}{RGB}{255,255,255}

\path[fill=fillColor] ( 39.72,244.27) rectangle ( 57.07,261.62);
\end{scope}
\begin{scope}
\definecolor{drawColor}{RGB}{152,78,163}

\path[draw=drawColor,line width= 1.1pt,line join=round] ( 41.46,252.94) -- ( 55.33,252.94);
\end{scope}
\begin{scope}
\definecolor{fillColor}{RGB}{255,255,255}

\path[fill=fillColor] (109.88,244.27) rectangle (127.22,261.62);
\end{scope}
\begin{scope}
\definecolor{drawColor}{RGB}{77,175,74}

\path[draw=drawColor,line width= 1.1pt,line join=round] (111.61,252.94) -- (125.49,252.94);
\end{scope}
\begin{scope}
\definecolor{fillColor}{RGB}{255,255,255}

\path[fill=fillColor] (198.56,244.27) rectangle (215.90,261.62);
\end{scope}
\begin{scope}
\definecolor{drawColor}{RGB}{55,126,184}

\path[draw=drawColor,line width= 1.1pt,line join=round] (200.29,252.94) -- (214.17,252.94);
\end{scope}
\begin{scope}
\definecolor{fillColor}{RGB}{255,255,255}

\path[fill=fillColor] (307.64,244.27) rectangle (324.98,261.62);
\end{scope}
\begin{scope}
\definecolor{drawColor}{RGB}{228,26,28}

\path[draw=drawColor,line width= 1.1pt,line join=round] (309.37,252.94) -- (323.25,252.94);
\end{scope}
\begin{scope}
\definecolor{drawColor}{RGB}{0,0,0}

\node[text=drawColor,anchor=base west,inner sep=0pt, outer sep=0pt, scale=  0.88] at ( 62.57,249.91) {\KaMinPar};
\end{scope}
\begin{scope}
\definecolor{drawColor}{RGB}{0,0,0}

\node[text=drawColor,anchor=base west,inner sep=0pt, outer sep=0pt, scale=  0.88] at (132.72,249.91) {Two-Phase LP};
\end{scope}
\begin{scope}
\definecolor{drawColor}{RGB}{0,0,0}

\node[text=drawColor,anchor=base west,inner sep=0pt, outer sep=0pt, scale=  0.88] at (221.40,249.91) {Graph Compression};
\end{scope}
\begin{scope}
\definecolor{drawColor}{RGB}{0,0,0}

\node[text=drawColor,anchor=base west,inner sep=0pt, outer sep=0pt, scale=  0.88] at (330.48,249.91) {One-Pass Contraction (\TeraPart)};
\end{scope}
\end{tikzpicture}
    \caption{
        Left and middle: Relative running times (all $k$ values) and peak memory ($k = \numprint{30000}$) on Benchmark Set B for \TeraPart relative to \KaMinPar when enabling the following optimizations one after the other: (i) two-phase label propagation, (ii) graph compression, and (iii) one-pass cluster contraction. 
        Edge cuts for \TeraPart are listed in \Cref{tbl:hwg-fm}.
        Right: Compression ratios with just gap encoding and gap encoding plus interval encoding.
    }
    \label{fig:hwg-time-memory}
\end{figure*}

\begin{table}
    \centering
    \caption{
        Edge cuts corresponding to \Cref{fig:hwg-time-memory} for $k = 64$.
        We report the edge cut for \TeraPart(-LP) as percentage of total edges cut and the edge cut of \TeraPart-FM relative to \TeraPart-LP.
    }
    \label{tbl:hwg-fm}
    \begin{tabular}{llrrrr}
        Graph & Algorithm & Cut & Time & Memory \\
        \midrule
        \multirow{2}{*}{\Instance{gsh-2015}} & \TeraPart-LP & 3.09\% & 46\,s & 68\,GiB \\
        & \TeraPart-FM & 0.92$\times$ & 355\,s & 125\,GiB \\
        \midrule
        \multirow{2}{*}{\Instance{clueweb12}} & \TeraPart-LP & 10.99\% & 71\,s & 71\,GiB \\
        & \TeraPart-FM & 0.94$\times$ & \numprint{1468}\,s & 170\,GiB \\
        \midrule
        \multirow{2}{*}{\Instance{uk-2014}} & \TeraPart-LP & 0.13\% & 56\,s & 55\,GiB \\
        & \TeraPart-FM & 0.95$\times$ & 65\,s & 102\,GiB \\
        \midrule
        \multirow{2}{*}{\Instance{eu-2015}} & \TeraPart-LP & 0.32\% & 77\,s & 87\,GiB \\
        & \TeraPart-FM & 0.96$\times$ & 196\,s & 158\,GiB \\
        \midrule
        \multirow{2}{*}{\Instance{hyperlink}} & \TeraPart-LP & 1.68\% & 191\,s & 279\,GiB \\
        & \TeraPart-FM & 0.87$\times$ & \numprint{2204}\,s & 538\,GiB \\
        \bottomrule
    \end{tabular}
\end{table}

Memory savings are arguably more important for huge graphs, and the graphs in Benchmark Set A are rather small compared to the memory available in modern high-end servers.
Thus, we now turn towards the much larger web graphs of Benchmark Set B (confer \Cref{table:hwg-stats}) to further showcase the scalability of \TeraPart.
Compression ratios (see \Cref{fig:hwg-time-memory}, right) for these graphs range from \RhwgMinCompressionRatioRounded{} for \Instance{hyperlink} to just over \RhwgMaxCompressionRatioRounded{} for \Instance{eu-2015}.
Interval encoding appears crucial for these graphs, as gap encoding alone only achieves compression ratios of \RhwgMinCompressionRatioGapOnly{} to \RhwgMaxCompressionRatioGapOnly{}.

In terms of running time, we generally observe the same pattern as before, with two-phase label propagation being the most impactful optimization. 
In particular, we observe a speedup of \RhwgEuSpVsTpTimeRatio{} for the \Instance{eu-2015} graph when enabling two-phase label propagation (which reduces to \RhwgEuSpVsUcTimeRatio{} after enabling graph compression and one-pass cluster contraction).
Graph compression generally increases running time, while one-pass contraction slightly decreases it.

\TeraPart only uses a fraction of the memory required by \KaMinPar, as shown in \Cref{fig:hwg-time-memory} (middle).
There, we highlight $k = \numprint{30000}$.
For smaller $k$, the graphs shrink faster and thus memory savings by one-pass contraction are less pronounced (\RhwgGcVsUcSmallKHeapPerc{}\% vs \RhwgGcVsUcLargeKHeapPerc{}\% reduction), while the influence of two-phase label propagation and graph compression remains similar.
On \Instance{gsh-2015}, \Instance{clueweb12}, \Instance{uk-2014} and \Instance{eu-2015}, \KaMinPar uses \RhwgGshKaMinParVsTeraPartHeapRatio{}, \RhwgCluewebKaMinParVsTeraPartHeapRatio{}, \RhwgUkKaMinParVsTeraPartHeapRatio{} and \RhwgEuKaMinParVsTeraPartHeapRatio{} times more memory than \TeraPart (averaged over all $k$), respectively.
For the largest graph in the benchmark set, \Instance{hyperlink}, \KaMinPar would require roughly 3.4\,TiB RAM, and thus runs out of memory on our 1.5\,TiB machine.
\TeraPart can partition this graph with just \RhwgHyperlinkMinHeapGiB\,GiB--\RhwgHyperlinkMaxHeapGiB\,GiB of memory, depending on $k$.

\begin{figure*}[t]
    \centering
    \input{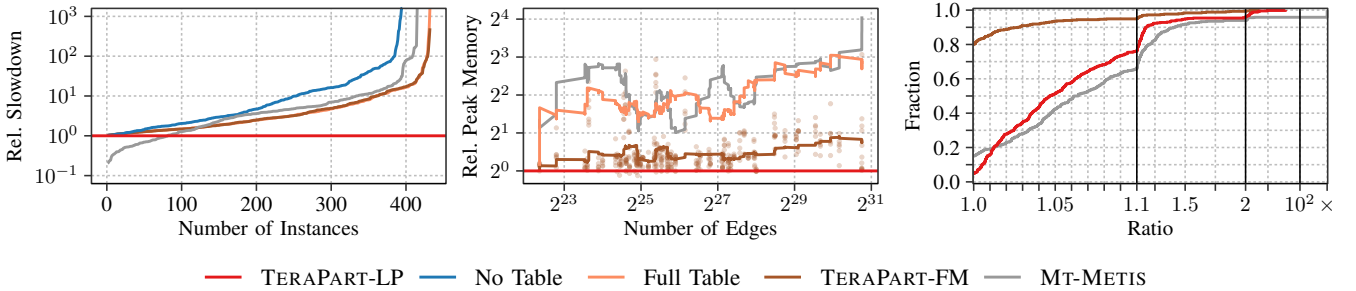}
\begin{tikzpicture}[x=1pt,y=1pt]
\definecolor{fillColor}{RGB}{255,255,255}
\begin{scope}
\definecolor{fillColor}{RGB}{255,255,255}

\path[fill=fillColor] (-54.89,238.77) rectangle (307.84,267.12);
\end{scope}
\begin{scope}
\definecolor{fillColor}{RGB}{255,255,255}

\path[fill=fillColor] (-49.39,244.27) rectangle (-32.05,261.62);
\end{scope}
\begin{scope}
\definecolor{drawColor}{RGB}{228,26,28}

\path[draw=drawColor,line width= 1.1pt,line join=round] (-47.66,252.94) -- (-33.78,252.94);
\end{scope}
\begin{scope}
\definecolor{fillColor}{RGB}{255,255,255}

\path[fill=fillColor] ( 31.75,244.27) rectangle ( 49.10,261.62);
\end{scope}
\begin{scope}
\definecolor{drawColor}{RGB}{31,120,180}

\path[draw=drawColor,line width= 1.1pt,line join=round] ( 33.49,252.94) -- ( 47.36,252.94);
\end{scope}
\begin{scope}
\definecolor{fillColor}{RGB}{255,255,255}

\path[fill=fillColor] ( 95.48,244.27) rectangle (112.83,261.62);
\end{scope}
\begin{scope}
\definecolor{drawColor}{RGB}{252,141,98}

\path[draw=drawColor,line width= 1.1pt,line join=round] ( 97.22,252.94) -- (111.09,252.94);
\end{scope}
\begin{scope}
\definecolor{fillColor}{RGB}{255,255,255}

\path[fill=fillColor] (162.22,244.27) rectangle (179.57,261.62);
\end{scope}
\begin{scope}
\definecolor{drawColor}{RGB}{166,86,40}

\path[draw=drawColor,line width= 1.1pt,line join=round] (163.96,252.94) -- (177.83,252.94);
\end{scope}
\begin{scope}
\definecolor{fillColor}{RGB}{255,255,255}

\path[fill=fillColor] (245.36,244.27) rectangle (262.71,261.62);
\end{scope}
\begin{scope}
\definecolor{drawColor}{gray}{0.60}

\path[draw=drawColor,line width= 1.1pt,line join=round] (247.10,252.94) -- (260.98,252.94);
\end{scope}
\begin{scope}
\definecolor{drawColor}{RGB}{0,0,0}

\node[text=drawColor,anchor=base west,inner sep=0pt, outer sep=0pt, scale=  0.88] at (-26.55,249.91) {\TeraPart-LP};
\end{scope}
\begin{scope}
\definecolor{drawColor}{RGB}{0,0,0}

\node[text=drawColor,anchor=base west,inner sep=0pt, outer sep=0pt, scale=  0.88] at ( 54.60,249.91) {No Table};
\end{scope}
\begin{scope}
\definecolor{drawColor}{RGB}{0,0,0}

\node[text=drawColor,anchor=base west,inner sep=0pt, outer sep=0pt, scale=  0.88] at (118.33,249.91) {Full Table};
\end{scope}
\begin{scope}
\definecolor{drawColor}{RGB}{0,0,0}

\node[text=drawColor,anchor=base west,inner sep=0pt, outer sep=0pt, scale=  0.88] at (185.07,249.91) {\TeraPart-FM};
\end{scope}
\begin{scope}
\definecolor{drawColor}{RGB}{0,0,0}

\node[text=drawColor,anchor=base west,inner sep=0pt, outer sep=0pt, scale=  0.88] at (268.21,249.91) {\MtMetis};
\end{scope}
\end{tikzpicture}
    \caption{
        Relative running times (left), peak memory (middle) and partition quality (right) for \TeraPart on Benchmark Set A, equipped with FM refinement using no gain cache (No Table), the full $O(nk)$ gain table (Full Table) and the proposed space-efficient gain table (\TeraPart-FM).
        For peak memory, we plot the per-instance ratios (only shown for \TeraPart-FM) with a right-aligned rolling geometric mean over 50 instances.
        As reference points, we include \TeraPart with label propagation refinement (\TeraPart-LP) and \MtMetis.
        We use $k \in \{8, 37, 64, 91, 128, \numprint{1000}\}$ since FM refinement for larger values of $k$ yields diminishing returns.
        Since all three FM configurations produce similar edge cuts, we only plot \TeraPart-FM in the performance profile (right).
    }
    \label{fig:ufm-fm}
\end{figure*}

\subsubsection{Scaling to a Trillion Edges}

To further explore the scalability of \TeraPart, we generate synthetic tera-scale \Instance{rgg2D} and \Instance{rhg} graphs using the aforementioned properties.
The generated \Instance{rgg2D} resp. \Instance{rhg} graphs have \numprint{8.59} billion vertices and \numprint{1.10} resp. \numprint{1.01} trillion (undirected) edges, which would require 16.1\,TiB resp. 14.8\,TiB to be stored as uncompressed CSR.
\TeraPart is able to compress these graphs down to just \numprint{1194}\,GiB resp. \numprint{608}\,GiB (compression ratio of \numprint{14.2} resp. \numprint{26.3}), allowing for in-memory partitioning on a single shared-memory machine.
Partitioning into $k = \numprint{30000}$ blocks takes just \numprint{663}\,s (\Instance{rgg2D}) resp. \numprint{467}\,s (\Instance{rhg}) while cutting \numprint{1.48}\% resp. \numprint{0.45}\% of the edges.
In total, \TeraPart requires 1.46\,TiB and 886\,GiB memory to partition the \Instance{rgg2D} and \Instance{rhg} graphs, respectively, allocating 304\,GiB and 278\,GiB of memory to auxiliary data structures.

\subsection{In-Memory with Parallel FM Refinement}

We now equip \TeraPart with FM refinement and evaluate the space-efficient gain tables introduced in \Cref{sec:fm}. 
We use \TeraPart-LP to refer to \TeraPart (i.e., with label propagation refinement) and \TeraPart-FM to refer to \TeraPart with additional FM refinement and and space-efficient gain tables.

On Benchmark Set A, partitioning with space efficient gain tables requires \RufmFmFullVsHashingHeapRatio{}$\times$ less memory than with the standard $O(nk)$ memory gain tables, compare \Cref{fig:ufm-fm}~(middle).
Perhaps surprisingly, these savings come at almost no cost in performance: on average, running time increases by just \RufmFmFullVsHashingTimePerc{}\%.
When we look only at graphs for which \TeraPart-FM uses more than 8\,GiB of memory, we find that our configuration using space-efficient gain tables reduces peak memory by factor \RufmFmFullVsHashingHeapRatioMedium{} compared to the standard gain table, while the penalty in running time diminishes to just \RufmFmFullVsHashingTimePercMedium{}\%.
For the largest graph (in terms of $n$) in Benchmark Set A, \Instance{kmer\_V1r}, the standard gain table runs out of memory for $k = \numprint{1000}$.
With space-efficient gain tables, we can successfully partition this graph with just \RufmFmKmerLargeKHeapGiB{}\,GiB of memory.

Using no gain table at all (i.e., gains are repeatedly recomputed from scratch instead) is not a viable option, as shown in \Cref{fig:ufm-fm}~(left).
On average, this configuration is \RufmFmHashingVsOnTheFlyTimeRatio{} times slower than \TeraPart-FM, with slowdowns of at least one order of magnitude on \RufmFmHashingVsOnTheFlyNumOrderOfMagnitudeSpeedup{} out of \RufmLpNumInstances{} instances.
It exceeds our one hour time limit on \RufmFmOnTheFlyTimeouts{} instances.

In terms of solution quality (\Cref{fig:ufm-fm} right), all three FM configurations produce similar edge cuts (thus we only plot one curve).
Unsurprisingly, \TeraPart-FM finds better cuts than \TeraPart-LP or \MtMetis on 80\% of the instances.
On 50\% of the instances, FM refinement reduces the edge cut by at least 4.5\% compared to just label propagation refinement.
While \MtMetis finds better cuts than \TeraPart-FM on roughly 15\% of the instances, we note once more that most of its partitions do not respect the balance constraint (\RufmLpMtMetisNumImbalancedInstances{} out of \RufmLpNumInstances{}).
Both \TeraPart-$\{$LP, FM$\}$ always produce balanced partitions.

With our space-efficient gain tables, FM refinement also scales to the much larger graphs of Benchmark Set B, see \Cref{tbl:hwg-fm}.
Here, edge cut improvements range from 4\% for \Instance{eu-2015} to 13\% for \Instance{hyperlink}.

\subsection{Distributed-Memory Partitioning}\label{sec:exp:horeka}

\begin{figure*}[t]
    \centering
    \input{figures/plots/horeka}
\begin{tikzpicture}[x=1pt,y=1pt]
\definecolor{fillColor}{RGB}{255,255,255}
\begin{scope}
\definecolor{drawColor}{RGB}{255,255,255}
\definecolor{fillColor}{RGB}{255,255,255}

\path[draw=drawColor,line width= 0.6pt,line join=round,line cap=round,fill=fillColor] ( 49.37,244.27) rectangle (305.43,261.62);
\end{scope}
\begin{scope}
\definecolor{fillColor}{RGB}{255,255,255}

\path[fill=fillColor] ( 49.37,244.27) rectangle ( 66.71,261.62);
\end{scope}
\begin{scope}
\definecolor{drawColor}{RGB}{217,95,2}

\path[draw=drawColor,line width= 0.6pt,line join=round] ( 51.10,252.94) -- ( 64.98,252.94);
\end{scope}
\begin{scope}
\definecolor{drawColor}{RGB}{217,95,2}
\definecolor{fillColor}{RGB}{217,95,2}

\path[draw=drawColor,line width= 0.4pt,line join=round,line cap=round,fill=fillColor] ( 58.04,252.94) circle (  1.96);
\end{scope}
\begin{scope}
\definecolor{fillColor}{RGB}{255,255,255}

\path[fill=fillColor] (119.34,244.27) rectangle (136.69,261.62);
\end{scope}
\begin{scope}
\definecolor{drawColor}{RGB}{117,112,179}

\path[draw=drawColor,line width= 0.6pt,line join=round] (121.08,252.94) -- (134.95,252.94);
\end{scope}
\begin{scope}
\definecolor{drawColor}{RGB}{117,112,179}
\definecolor{fillColor}{RGB}{117,112,179}

\path[draw=drawColor,line width= 0.4pt,line join=round,line cap=round,fill=fillColor] (128.01,252.94) circle (  1.96);
\end{scope}
\begin{scope}
\definecolor{fillColor}{RGB}{255,255,255}

\path[fill=fillColor] (181.28,244.27) rectangle (198.62,261.62);
\end{scope}
\begin{scope}
\definecolor{drawColor}{RGB}{27,158,119}

\path[draw=drawColor,line width= 0.6pt,line join=round] (183.01,252.94) -- (196.89,252.94);
\end{scope}
\begin{scope}
\definecolor{drawColor}{RGB}{27,158,119}
\definecolor{fillColor}{RGB}{27,158,119}

\path[draw=drawColor,line width= 0.4pt,line join=round,line cap=round,fill=fillColor] (189.95,252.94) circle (  1.96);
\end{scope}
\begin{scope}
\definecolor{fillColor}{RGB}{255,255,255}

\path[fill=fillColor] (246.06,244.27) rectangle (263.41,261.62);
\end{scope}
\begin{scope}
\definecolor{drawColor}{RGB}{231,41,138}

\path[draw=drawColor,line width= 0.6pt,line join=round] (247.80,252.94) -- (261.67,252.94);
\end{scope}
\begin{scope}
\definecolor{drawColor}{RGB}{231,41,138}
\definecolor{fillColor}{RGB}{231,41,138}

\path[draw=drawColor,line width= 0.4pt,line join=round,line cap=round,fill=fillColor] (254.73,252.94) circle (  1.96);
\end{scope}
\begin{scope}
\definecolor{drawColor}{RGB}{0,0,0}

\node[text=drawColor,anchor=base west,inner sep=0pt, outer sep=0pt, scale=  0.80] at ( 72.21,250.19) {\dKaMinPar};
\end{scope}
\begin{scope}
\definecolor{drawColor}{RGB}{0,0,0}

\node[text=drawColor,anchor=base west,inner sep=0pt, outer sep=0pt, scale=  0.80] at (142.19,250.19) {\ParMetis};
\end{scope}
\begin{scope}
\definecolor{drawColor}{RGB}{0,0,0}

\node[text=drawColor,anchor=base west,inner sep=0pt, outer sep=0pt, scale=  0.80] at (204.12,250.19) {\xTeraPart};
\end{scope}
\begin{scope}
\definecolor{drawColor}{RGB}{0,0,0}

\node[text=drawColor,anchor=base west,inner sep=0pt, outer sep=0pt, scale=  0.80] at (268.91,250.19) {\XtraPuLP};
\end{scope}
\begin{scope}
\definecolor{drawColor}{RGB}{255,255,255}
\definecolor{fillColor}{RGB}{255,255,255}

\path[draw=drawColor,line width= 0.6pt,line join=round,line cap=round,fill=fillColor] (333.88,244.27) rectangle (456.52,261.62);
\end{scope}
\begin{scope}
\definecolor{fillColor}{RGB}{255,255,255}

\path[fill=fillColor] (333.88,244.27) rectangle (351.22,261.62);
\end{scope}
\begin{scope}
\definecolor{fillColor}{RGB}{0,0,0}

\path[fill=fillColor] (342.55,252.94) circle (  1.96);
\end{scope}
\begin{scope}
\definecolor{fillColor}{RGB}{255,255,255}

\path[fill=fillColor] (392.17,244.27) rectangle (409.51,261.62);
\end{scope}
\begin{scope}
\definecolor{drawColor}{RGB}{0,0,0}

\path[draw=drawColor,line width= 0.4pt,line join=round,line cap=round] (400.84,256.00) --
	(403.48,251.42) --
	(398.20,251.42) --
	cycle;
\end{scope}
\begin{scope}
\definecolor{drawColor}{RGB}{0,0,0}

\node[text=drawColor,anchor=base west,inner sep=0pt, outer sep=0pt, scale=  0.80] at (356.72,250.19) {Feasible};
\end{scope}
\begin{scope}
\definecolor{drawColor}{RGB}{0,0,0}

\node[text=drawColor,anchor=base west,inner sep=0pt, outer sep=0pt, scale=  0.80] at (415.01,250.19) {Imbalanced};
\end{scope}
\end{tikzpicture}
    \caption{
        Results obtained on the \HoreKa cluster.
        Left and middle: Comparison of \xTeraPart against \ParMetis and \XtraPuLP on $8$ compute nodes with increasing \Instance{rgg2D} and \Instance{rhg} graphs.
        \XtraPuLP runs out of memory starting at $2^{35}$ edges on both families, while \ParMetis runs out of memory on \Instance{rgg2D} graphs and exceeds our 1\,h time limit on \Instance{rhg} graphs.
        Edge cuts are compared in \Cref{tbl:horeka}.
        Right: Weak scaling results on \Instance{rgg2D} and \Instance{rhg} graphs for \xTeraPart with the largest feasible \Instance{rgg2D} and \Instance{rhg} graphs up to 128 compute nodes.
        The largest graphs have $2^{37}$ vertices and $2^{44}$ edges, which \xTeraPart can partition less than 10 minutes.
    }
    \label{fig:horeka}
\end{figure*}

Lastly, we equip the distributed-memory version of \KaMinPar (i.e., \dKaMinPar) with the same compression techniques as implemented in \TeraPart and obtain \xTeraPart.
Here, we focus on generated \Instance{rgg2D} and \Instance{rhg} graphs, generally use $3$ repetitions, $k = 64$ and $\varepsilon = 3\%$.

\Cref{fig:horeka}~(left and middle) compares \xTeraPart against \dKaMinPar, \ParMetis~\cite{PARMETIS} and \XtraPuLP~\cite{xtrapulp} by partitioning graphs of growing sizes while keeping the number of compute nodes fixed to $8$ (i.e., with 2\,TiB RAM in total). 
We obverse that \xTeraPart is able to partition graphs of both families with up to $2^{40}$ edges, closing the line to our single-machine experimentation.
Since distributed graph partitioning requires additional auxiliary data structures (refer \Cref{sec:dkaminpar}), \xTeraPart requires slightly more memory to partition these graphs than \TeraPart (1.2$\times$--\,1.3$\times$).
\emph{Both} \ParMetis and \XtraPuLP only manage to partition graphs of either family that are $64$ times smaller than that, running out of memory or exceeding our one hour time limit already at $2^{35}$ edges. 
Without graph compression, \dKaMinPar can partition graphs with up to $2^{37}$ edges, i.e., $8$ times smaller than \xTeraPart, requiring 4.8$\times$--\,4.5$\times$ more memory. 
Compared to the medium-sized graphs of Benchmark Set A, we observe that the compression techniques introduce higher running time penalties for these graphs.

In \Cref{fig:horeka}~(right), we show weak scaling results up to 128 compute nodes (the largest number of nodes that the cluster can allocate to a single job and that is also a power of two).
We observe good weak scaling behavior for both graph families. 
On 128 compute nodes, \xTeraPart can partition graphs with $2^{44}$ edges in slightly less than 10 minutes.

\begin{table}
    \centering
    \caption{
        Edge cuts corresponding to \Cref{fig:horeka} (left and middle), reported relative to $m$ (\xTeraPart) resp. \xTeraPart (\ParMetis, \XtraPuLP).
        Imbalanced partitions are marked with *, out-of-memory with OOM and out-of-time with OOT ($> 1$\,h).
    }
    \label{tbl:horeka}
    \addtolength{\tabcolsep}{-0.1em}
    \begin{tabular}{l|lrrrr}
        \multicolumn{1}{c}{}                              & \multicolumn{1}{c}{}          & \multicolumn{4}{c}{Cut rel. to $m$ resp. \xTeraPart}                                            \\
        \cmidrule{3-6}
        \multicolumn{1}{c}{$G$}                           & \multicolumn{1}{l}{Algorithm} & $m = 2^{32}$                            & $m = 2^{33}$      & $m = 2^{34}$      & $m = 2^{35}$ \\
        \midrule
        \multirow{3}{*}{\rotatebox{90}{\Instance{rgg2D}}} & \xTeraPart                    & 0.93\%                              & 0.67\%        & 0.48\%        & 0.35\%   \\
        & \ParMetis                     & 1.00$\times$                        & 1.00$\times$  & 0.99$\times$  & OOM      \\
        & \XtraPuLP                     & *5.56$\times$                       & *5.56$\times$ & *6.18$\times$ & OOM      \\
        \midrule
        \multirow{3}{*}{\rotatebox{90}{\Instance{rhg}}}   & \xTeraPart                    & 0.23\%                              & 0.16\%        & 0.11\%        & 0.04\%   \\
        & \ParMetis                     & 1.12$\times$                        & 1.14$\times$  & 0.86$\times$           & OOT      \\
        & \XtraPuLP                     & 48.40$\times$                       & 57.63$\times$ & 68.44$\times$ & OOM      \\
        \bottomrule
    \end{tabular}
\end{table}

\section{Alternative Approaches}

Besides optimizing the multilevel framework for memory efficiency as done in this paper, there are two existing approaches that reduce the burden on RAM: semi-external memory algorithms and streaming algorithms. 
These approaches do not hold the graph in memory. 
Instead, the vertex neighborhoods are loaded one at a time from network or SSD, processed and then dropped again. 
As streaming algorithms perform just one pass over the graph, it is well known that they achieve sub-par solution quality compared to multilevel algorithms. 
We found that \HeiStream~\cite{HEISTREAM} (which produces lower edge cuts than more basic streaming algorithms) cuts \RtegHeiStreamVsTeraPartRggCutRatio{}$\times$ (\Instance{rgg2D}) to \RtegHeiStreamVsTeraPartRhgCutRatio{}$\times$ (\Instance{rhg}) more edges than \TeraPart on our generated tera-edge graphs for $k = \numprint{30000}$.

\begin{table}
    \centering
    \caption{
        Comparing \TeraPart against the semi-external memory partitioning algorithm (\SEM) from~\cite{ASS15} with $k=16$ and $\varepsilon = 3\%$.
    }
    \begin{tabular}{llrrr}
        Graph                                   & Algorithm & Cut [M] & Time [s] & Memory [GiB] \\
        \midrule
        \multirow{2}{*}{\Instance{arabic-2005}} & \TeraPart & 1.88    & 5.5      & 1.44         \\
        & \SEM      & 2.28    & 36.0     & 1.85         \\
        \midrule
        \multirow{2}{*}{\Instance{uk-2002}}     & \TeraPart & 1.45    & 3.2      & 1.05         \\
        & \SEM      & 1.54    & 39.7     & 1.53         \\
        \midrule
        \multirow{2}{*}{\Instance{sk-2005}}     & \TeraPart & 15.68   & 18.3     & 3.62         \\
        & \SEM      & 22.25   & 203.4    & 7.76         \\
        \midrule
        \multirow{2}{*}{\Instance{uk-2007}}     & \TeraPart & 4.09    & 27.8     & 6.81         \\
        & \SEM      & 4.55    & 209.1    & 8.58         \\
        \bottomrule
    \end{tabular}
    \label{tbl:shm:vsexternal}
\end{table}

Semi-external memory algorithms perform multiple passes over the graph and can use $O(n)$ space for
auxiliary data. 
Thus, it is possible to implement label propagation and the multilevel framework in semi-external memory~\cite{ASS15}, whereas sophisticated heuristics such as FM~\cite{FM} seem difficult.
We are aware of one semi-external memory algorithm by Akhremtsev \etal\cite{ASS15}. This algorithm is an order of magnitude slower than \TeraPart, see \Cref{tbl:shm:vsexternal}. 
As their source code is not available, we compare against results on four graphs from their paper. 
To enable a fair comparison, we benchmarked \TeraPart on an equivalent machine hosting two 8-core Intel Xeon E5-2650v2 clocked at 2.6\,GHz.
We also note that semi-external algorithms are orthogonal to our work, and combining the two approaches is mutually beneficial to enable processing even larger graphs, though we leave this direction for future work.

\section{Conclusion}

We presented \TeraPart, the first multilevel algorithm capable of partitioning trillion-edge graphs on a single multi-core machine.
Showcasing \TeraPart's excellent efficiency, it does so in just under 8 minutes.
Moreover, its distributed version \xTeraPart can partition 16-trillion edge graphs in about 10 minutes.
\TeraPart consists of a collection of techniques to reduce the memory consumption of the coarsening and uncoarsening stages of the multilevel framework.
By splitting label propagation into two phases, we are able to reduce the auxiliary memory used from $O(np)$ down to $O(n)$, and simultaneously alleviate a previous load balancing bottleneck.
We apply the same two-phase technique to the graph contraction algorithm, and eliminate a copy of the coarse graph that was necessary due to the previous parallelization scheme.
With our space-efficient gain table, we reduce the memory consumption for FM refinement from $O(nk)$ to $O(m)$, resulting in a 5.8$\times$ peak memory reduction for graphs over 8\,GiB.
The last piece of the puzzle is storing the graph in compressed format and decoding neighborhoods on-the-fly when they are requested.
Based on our experiments, the next avenue for optimization is to reduce the memory for the input graph.
We are both interested in improving the compression ratios, as well as integrating our techniques with a semi-external algorithm.

 
\ifsubmission
\else
\section*{Acknowledgments}
This work was performed on the HoreKa supercomputer funded by the Ministry of Science, Research and the Arts Baden-Württemberg and by the Federal Ministry of Education and Research.
This project has received funding from the European Research Council (ERC) under the European Union’s Horizon 2020 research and innovation program (grant agreement No. 882500).
\begin{center}\includegraphics[scale=0.5]{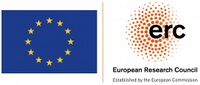}\end{center}
\fi
 
 \bibliographystyle{IEEEtran}
 \bibliography{references}

 \ifsubmission
 \else
 \clearpage
\appendix 

\noindent\begin{minipage}{\textwidth}
    \centering
    \input{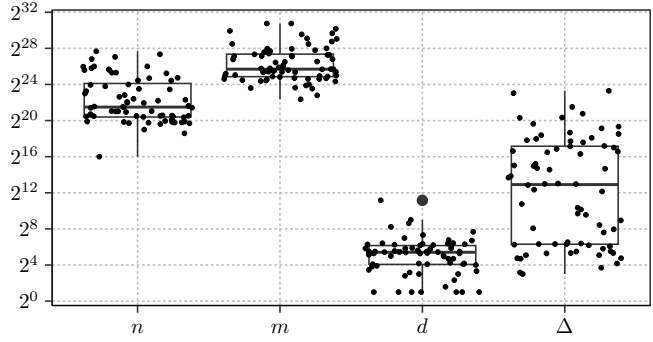}
    \captionof{figure}{
        Basic properties of the graphs in Benchmark Set A: number of vertices $n$, number of undirected edges $m$, average degree $d$ and maximum degree $\Delta$.
    }
    \label{fig:benchmark-set-a}
\end{minipage}

\noindent\begin{minipage}{\textwidth}
    \centering
    \input{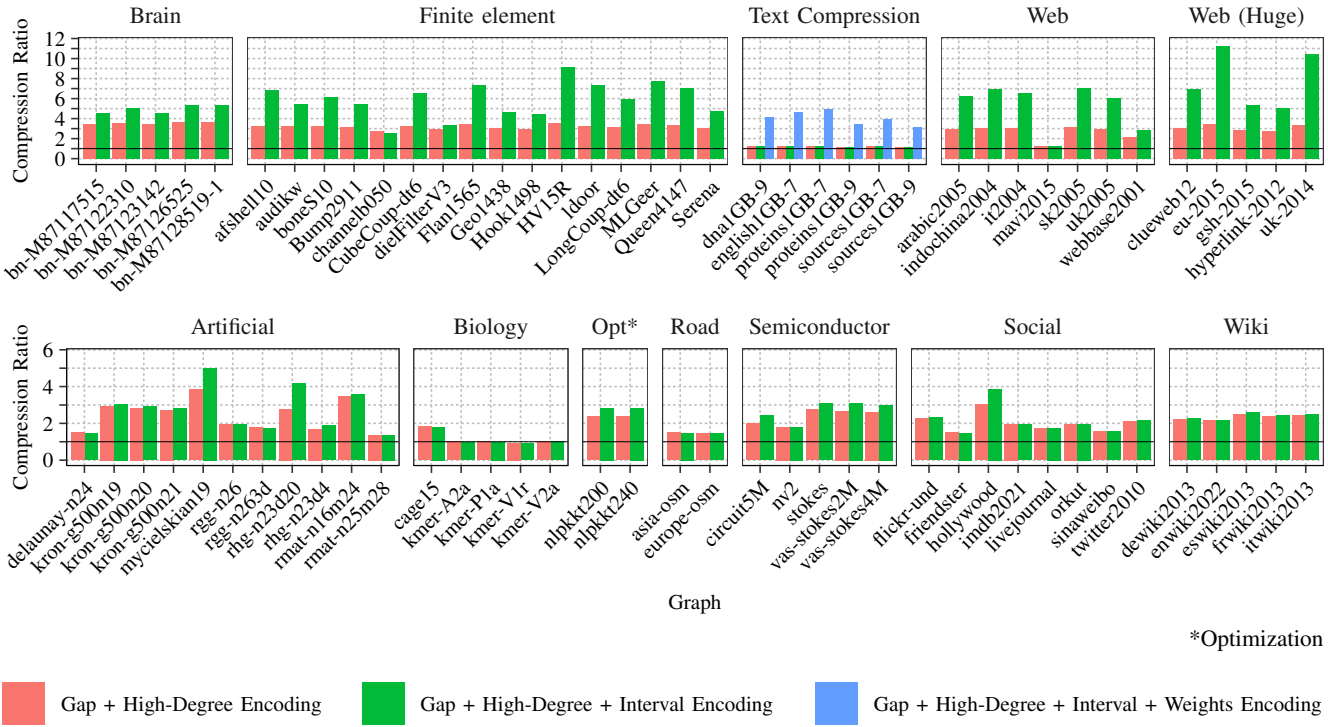}
    \captionof{figure}{
        Compression ratios of all graphs in Benchmark Set A and B.
        Improved compression ratios due to edge weight compression are only shown for graphs with non-uniform edge weights (class \emph{Text Compression} only).
    }
    \label{fig:compression-ratios}
\end{minipage}

 \fi

\end{document}